\documentclass[a4paper,11pt]{article}
\pdfoutput=1 

\usepackage{jcappub} 

\usepackage[T1]{fontenc} 

\title{\boldmath Measurement of atmospheric muon angular distribution using a portable setup of liquid scintillator bars}

\author{Hariom Sogarwal}
\author{and Prashant Shukla}
\affiliation{Nuclear Physics Division, Bhabha Atomic Research Centre, Mumbai 400085, India, } 
\affiliation{Homi Bhabha National Institute, Anushakti Nagar, Mumbai 400094, India}

\emailAdd{sogarwalhariom@gmail.com}
\emailAdd{pshuklabarc@gmail.com}

\abstract {Measurements of cosmogenic particles at various locations and altitudes are
becoming increasingly important in view of worldwide interests in rare signals for search of new
physics.
In this work, we report measurement of muon zenith angle distributions and integrated
flux using a portable setup of four one-meter long liquid scintillator bars.
Each scintillator bar is read out from both sides via photomultiplier tubes followed
by an 8-channel Digitizer. We exploit energy deposition and excellent timing of scintillators to construct two
dimensional tracks and hence angles of charged particles. We use liquid scintillators since they
have an added advantage of pulse shape discrimination (PSD) which can be used for detecting
muon induced particles. 
 The energy deposition, time window of event and PSD cuts are used to reduce the random as well as
correlated backgrounds. In addition, we propose three track quality parameters which are applied
to obtain a clean muon spectrum.
The zenith angle measurement is performed upto $60^\circ$.
With our improved analysis, we demonstrate that a setup of 3 bars can also be 
used for quicker and precise measurements. 
 The vertical muon flux measured is $66.70 \pm 0.36 \pm 1.50$ $/m^2/sr/s$ with $n=2.10 \pm 0.05 \pm 0.25 $
in $\cos^n \theta$ at the location of Mumbai, India ($19^{\circ}$N, $72.9^{\circ}$E) at Sea 
level with a muon momentum above $255$ MeV/$c$. 
The muon flux has dependence on various factors, the most prominents are latitudes, altitudes 
and momentum cut of muon so that portable setup like this can be a boon for such 
measurements at various locations.
}

\begin{document}

\maketitle
\flushbottom

\section{Introduction}
 The cosmic particles and their interactions with the Earth continues to be
an active area of physics for many decades~\cite{PDG2021}. 
The precise measurements of cosmogenic particles at various locations and altitudes
not only contains rich physics but is important in 
the context of searching for rare signals to answer many of the open problems of physics.
The cosmic particles are produced in cosmological processes like
Supernova, active galactic nuclei, pulsars, gamma ray bursts, the Sun and Big bang 
and are accelerated in an intergalactic field. 
The primary cosmic rays bombarding the Earth isotropically contain
high energy protons, helium and a tiny fraction of higher-Z nuclei.
The magnetic field of the Sun tends to exclude lower energy particles $E<1$ GeV
and during periods of high solar activity, less cosmic rays can reach the Earth.
Earth's magnetic field also tends to exclude lower energy charged particles and 
thus their flux is more towards poles as compared to that at equator.

  After entering the Earth atmosphere, the primary charged particles interact strongly with the
nuclei of air molecules (mainly oxygen and nitrogen) in the upper part of atmosphere
and produce showers of secondary particles containing mostly pions and kaons plus other particles 
like hyperons, charmed particles and nucleon-antinucleon pairs~\cite{Grieder2001,Rossi:1948fwx}.
The neutral pions decay to two photons and the charged pions and kaons mostly decay to
muons and neutrinos.
Thus, the bulk of particles reaching the Earth surface contains neutrinos and muons.
Although both
are closely linked, the former are least interacting and the later undergo Coulomb
interactions with the matter. 
 The muons are the most abundant energetic charged particles with a mean energy of about 
2 GeV at Sea level and are of immense use in testing and calibrating detectors 
for experimental high energy physics. 
 The muon zenith angle ($\theta$) distribution at Earth is observed to follow 
a distribution $I(\theta) = I_{\circ} \cos^n \theta$, where 
$I_{\circ}$ is the vertical muon flux and the exponent $n$ is $\sim 2$  \cite{Shukla2018}.
The parameters $I_{\circ}$ and $n$ depend on momentum cut off, latitude and altitude.
Measurements of muon flux have been performed at many locations with varying altitudes
right from the time cosmic rays were discovered.

Some of the early measurements of muon flux in the world are 
Hayman et al. (Durham, UK)~\cite{Hayman1962}, Greisen~\cite{Greisen1942}, 
Judge and Nash~\cite{Judge1965} and Crookes and Rastin~\cite{Crookes:1972xd}. 
Barbouti and Rastin~\cite{Grieder2001} measured the muon flux at Nottingham, UK using
flash tubes and Gieger counters in 1983. 
Haino et al. measured the energy distribution of muons at Tsukuba, Japan~\cite{Haino2004}.
Fukui et al.~\cite{Fukui1957} performed the measurements at geomagnetic latitude $24^\circ$N.
Allkofer et al.~\cite{Allkofer1968} measured the muon flux at latitude $9^\circ$N. 
The charge ratio of atmospheric muons is of considerable physics interest and there are
recent measurements like Agafonova et al.~\cite{OPERA:2010cos}.

 In India, the measurements of muon flux have been done at few places. 
Gokhale et al.~\cite{Grieder2001} in 1993 measured the muon flux in
Delhi ($28.7^\circ$N) using GM counter telescope.
In 1973, Karmakar et al.~\cite{Karmakar1973} measured the flux in Darjeeling ($16^\circ$N)
using plastic scintillator paddles. 
Sinha and Basu~\cite{Sinha1959} measured muon flux using a multiplate cloud chamber in Kolkata 
($22.57^\circ$N) in 1959. 
De et al.~\cite{De:1972nv} measured muon flux using an array of Gieger counters and
scintillators in Kolkata ($22.57^\circ$N) in 1972. 
Recently, Pethuraj et al.~\cite{pethuraj2017} performed the measurement using
a setup of 12 layers of Resistive Plate Chambers (RPCs) at Madurai which is close
to the proposed site for India based Neutrino Observatory (INO).
Pal et al.~\cite{pal2012} measured the muon flux in Colaba, Mumbai using a setup
of 12 layers of RPCs.  
 A country like India covers a wide range of latitudes ranging from
Srinagar ($34.08^{\circ}$N, $74.80^{\circ}$E) to Kanyakumari ($8.09^{\circ}$N, $77.53^{\circ}$E)
and this requires systematic measurements within India itself.
 In addition, many of the muon measurements are old and with the development of newer techniques
in detectors and electronics, a revisit to the measurements is warranted.

 In recent times, a lot of experiments have been done using plastic scintillators paddles/bars to
measure muon flux.
Wu et. al ~\cite{Wu:2013cno} used plastic scintillators to measure muon flux at
various underground depths in China. Recently, four plastic scintillators were used to
measure the muon flux at an underground laboratory at Jaduguda, India \cite{sharan2021}.
 Work in Ref.~\cite{blackmore2015} uses large area plastic scintillator and gives the energy
spectra of passing and stopping muons and decay electrons/positrons at ground and
at depth.
There are several recent works using plastic scintillators to measure muon angle
distributions using plastic scintillator bars
\cite{Lin:2010zze, Bahmanabadi:2019rro}.
 But setups using both energy and timing properties of scintillator bars for
constructing clean muon tracks and effectively using them for zenith angle distributions
are scarce.
 The work in Ref.~\cite{KuoThesisMIT2010} constructs muon tracks using two
scintillator bars read out from both ends.
The position in each bar was obtained using time difference between two ends of
the bar, albeit with a poor position resolution. 
We employ liquid scintillator bars to measure the muons. This is done with a
future aim of using the same setup for muon induced neutrons and spallation
particles exploiting the pulse shape descrimination (PSD) property of liquid
scintillators. To measure muon signal, a lower energy threshold is applied on
all scintillators which removes bulk of the background and an additional cut
on PSD improves the signal to background slightly more. Liquid scintillators
have the advantage that they offer uniformity of scintillating solutes in the
organic liquid. One practical drawback of using the liquid scintillators is
horizontal level across the length of the bar is to be maintained. The rise
time of liquid scintillators is slightly higher than the plastic scintillators.

 In this work, we perform measurement of muon flux and angular distributions using a portable
setup of four one-meter long liquid scintillator bars.
Each scintillator bar is read out from both sides via photomultiplier tubes followed
by an 8-channel Digitizer. 
We implemented several innovative ideas in the measurement.
 Position calibration of scintillator bars is performed using vertical cosmic muons constrained
by placing the four scintillators in cross positions. We also explore less time consuming
methods to connect the position with the measured time. 
Energy calibration of detectors has been done using muons tracks with different angles
travelling different pathlengths in the detector.
The energy deposition, time window of event and PSD cuts are used to reject the backgrounds.
In addition, three track quality parameters are proposed which are applied
to obtain a clean muon spectrum.
We optimize the distances between the bars so as to have larger zenith angle coverage
keeping a good angular resolution.
The setup can distinguish downgoing and upgoing tracks. 
 The acceptance of all the geometries is obtained by Monte Carlo simulations using
inputs from measured energy distributions. The efficiencies of all the cuts
are measured using experimental methods.
The correction in 2D track angles due to finite $\phi$ coverage has been done via simulation.  
 The setup can easily be used for studying the interaction of muons with materials
and can detect produced photons, protons and neutrons.

  The paper is organized as follows. After an Introduction, Section 2 describes the
experimental setups. Section 3 describes analysis methods in detail which
includes calibration of different parameters and track reconstruction.
Section 4 presents the methods to obtain the efficiencies of different cuts. 
Section 5 gives MC simulation used
in the analysis. The measurements via various geometries are described in Section 6
and the results with systematic and statistical uncertainties are described in Section 7.
Finally, we summarize in section 8.

\section{Experimental setups}
To measure the muon flux at sea level, a setup of four liquid scintillator bars 
placed in the same vertical plane is used. Each detector bar is 1.02 meter long
and with a cross section 4.9 cm (height) $\times$ 5.6 cm (width) of active material.
Each bar consists of Organic liquid scintillator EJ301 (density 0.874 gm/cm$^{3}$) 
enclosed inside a cuboidal aluminum casing with glass window at each end which were
coupled to the photomultiplier tubes using optical gel.
The PMTs used are from Electron Tubes Enterprises Ltd with
model number 9214KA which have diameter 51 mm with blue-green sensitive
bialkali photocathode and 12 high gain high stability dynodes
of linear focused design.
We use the energy, time and shape of the pulse in the measurements.  
Anode of PMT acquires an amplified negative signal which is fed to the attenuator and
then to Digitizer DT5730.
The CoMPASS Software is used to store the integrated energy
in Long and Short gates and a time stamp for signals in each individual channel.
Later we use an offline algorithm to reconstruct and analyze muon tracks.
Here 25 ns time width is taken for searching the events 
in 8 signals. Later, after the time calibration we constrain 
the time difference of events between bottom and top detectors within
7 ns. For Close geometries we reduce this window to 5 ns.

\begin{figure}
\begin{center}
\includegraphics[width=12cm, height=6cm]{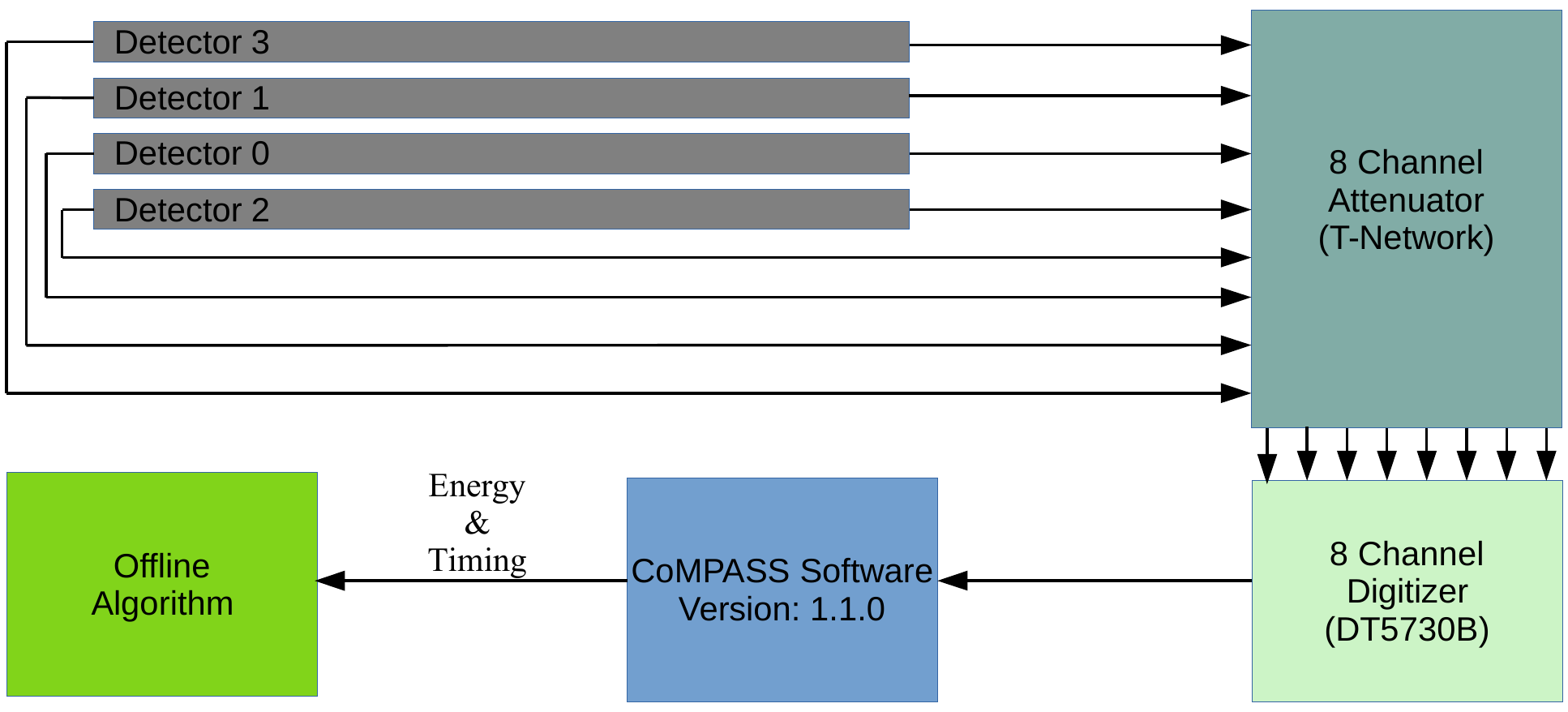}
\caption{Block diagram of the experimental setup with the data acquisition scheme.}
\label{fig1block}
\end{center}
\end{figure}

\begin{figure}
\begin{center}
\includegraphics[width=11cm, height=11cm]{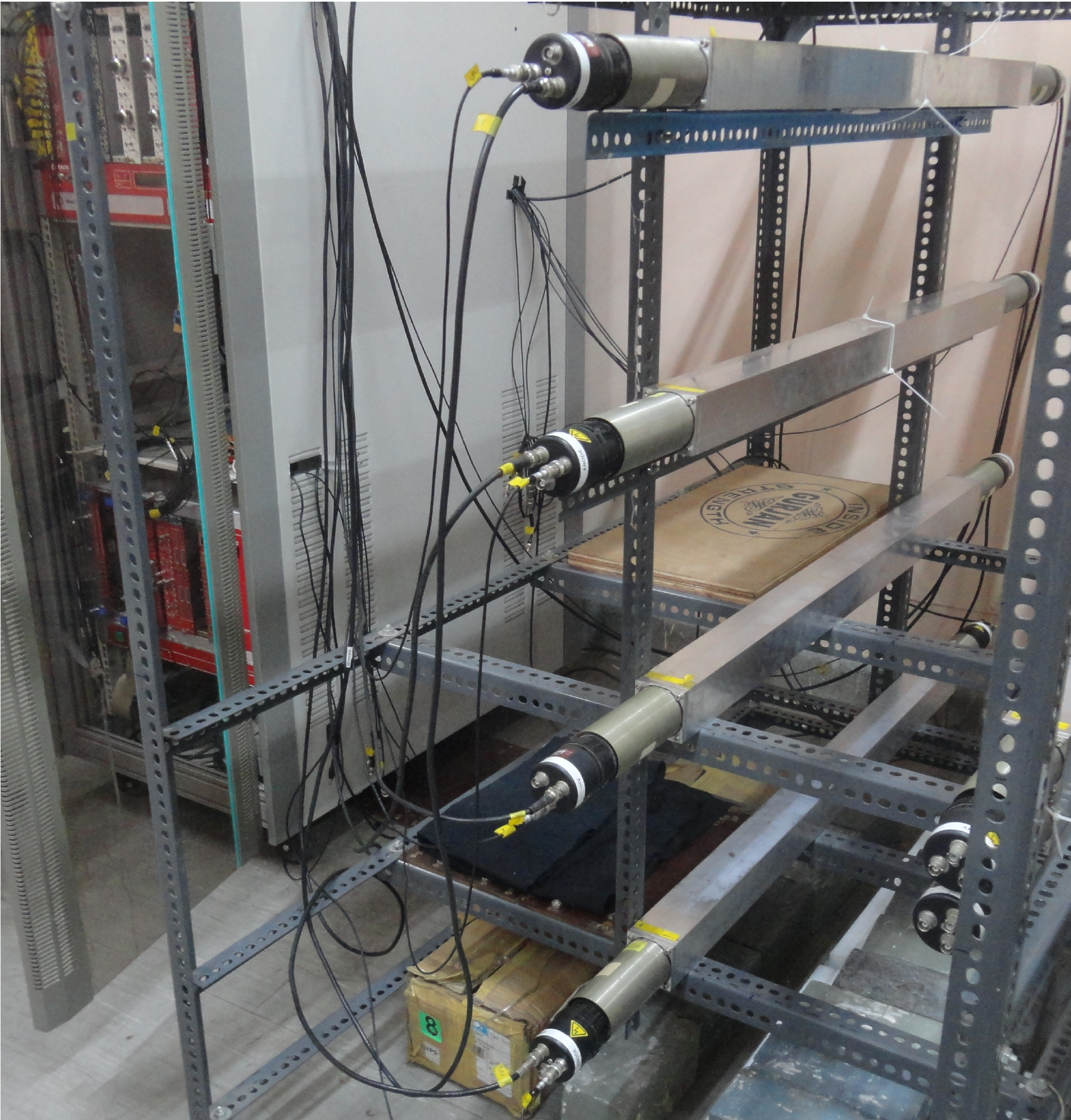}
\caption{Experimental setup of 4 liquid scintillator bars in Square geometry.}
\label{fig2det}
\end{center}
\end{figure}

\begin{figure}
\begin{center}
\includegraphics[width=7.5cm, height=5cm]{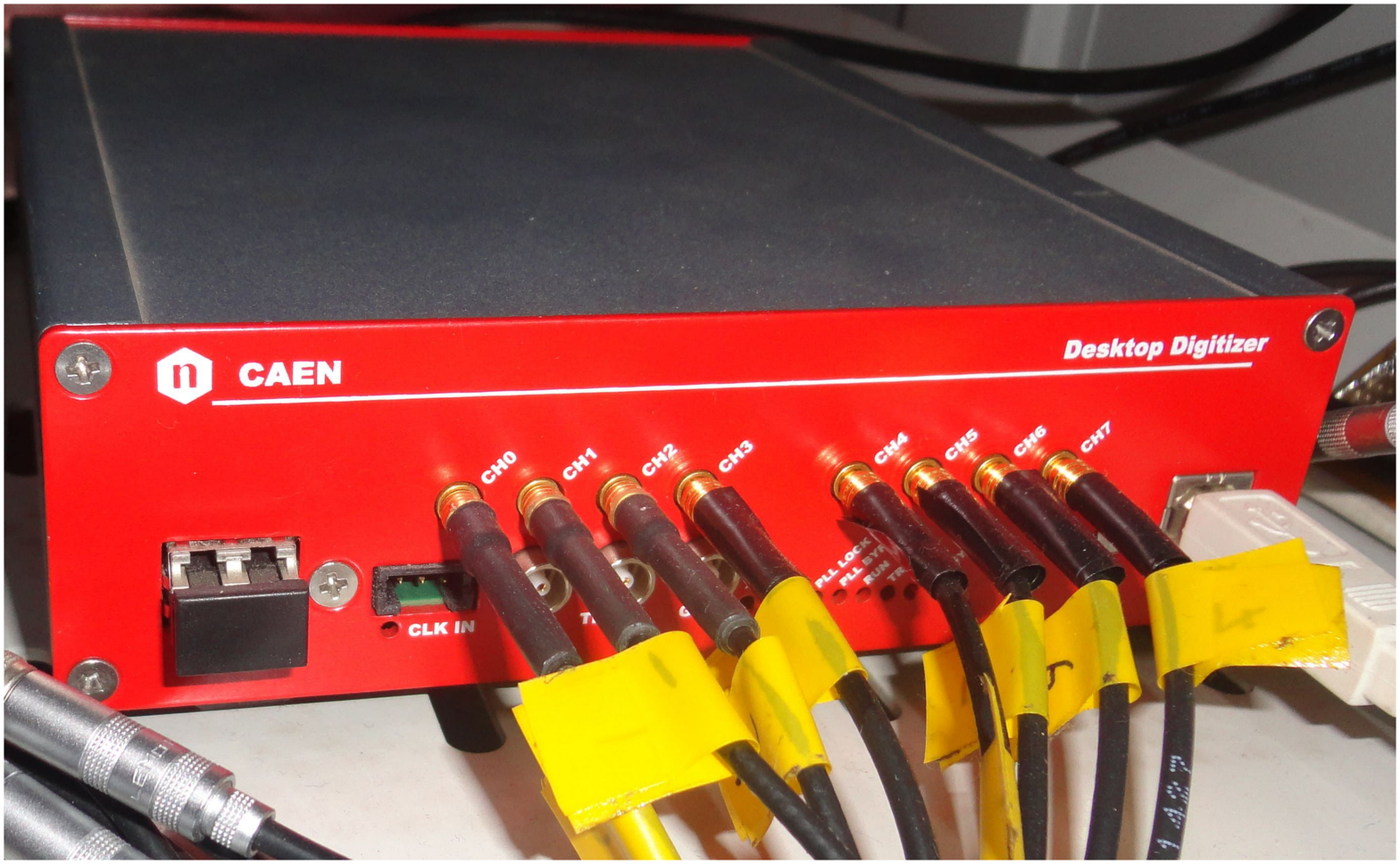}
\includegraphics[width=7.5cm, height=5cm]{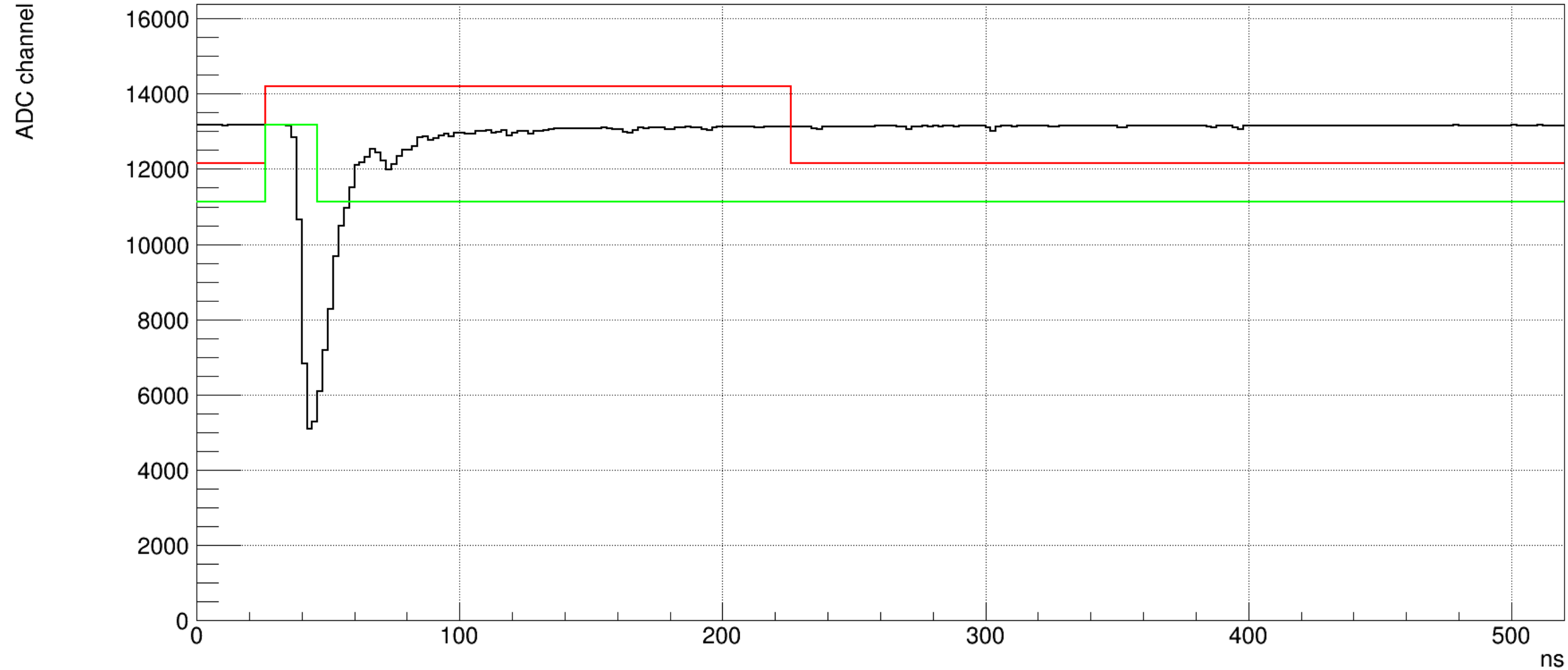}
\caption{Picture of 8-Channel Digitizer (DT5730) along with a typical digitized pulse (ADC channel number as a function of time (ns).) from liquid scintillator.}
\label{fig3pulse}
\end{center}
\end{figure}

 Figure~\ref{fig1block} shows the block diagram of the experimental setup 
with the data acquisition scheme.
Figure~\ref{fig2det} shows the experimental setup of 4 liquid scintillator bars in Square
geometry.
Figure~\ref{fig3pulse} shows the picture of 8-Channel Digitizer (DT5730) along with a
typical digitized pulse (ADC channel number as a function of time (ns).) from liquid scintillator.
The Long and Short gates timings are also shown which will be used to integrate the charge of 
pulse within their time width. The time widths of Long and Short gates for charge integration 
are given in Table~\ref{table1} and the discriminator parameters are given in Table~\ref{table2}.
The value of Short gate is taken as 20 ns and is changed to 24 ns for Small and
Close geometries. The acquisition rate in each channel $\approx$ 80 Hz.

\begin{table}[!htb]
\begin{minipage}{.5\linewidth}
\caption{Digitizer QDC Section.}
\label{table1}
\centering
\begin{tabular}{|c|c|} 
\hline
{\bf Parameter} & {\bf Values} \\
\hline
{\bf Energy coarse gain} & 10fC/(LSB$\times$V$_{pp}$) \\ 
{\bf Gate} & 200 ns\\
{\bf Short gate} & 20/24 ns\\
{\bf Pre-gate} & 70 ns\\
\hline
\end{tabular}
\end{minipage}%
\begin{minipage}{.5\linewidth}
\caption{Digitizer Discriminator Section.}
\label{table2}
\centering
\begin{tabular}{|c|c|} 
\hline
{\bf Parameter} & {\bf Values} \\ 
\hline
{\bf Discriminator mode} & CFD \\ 
{\bf Threshold} & 400 lsb\\
{\bf Trigger holdoff} & 1024 ns\\
{\bf CFD delay} & 6 ns\\
{\bf CFD fraction} &  75$\%$\\
{\bf Input Smoothing} &  2 samples\\
\hline
\end{tabular}
\end{minipage} 
\end{table}

We have used many geometries of the four detector bars in our setup.
We first assign numbers 0, 1, 2, and 3 to the four bars used in the setup to explain 
each geometry.
In all geometries, the sequence of detectors arranged from top to bottom is (3,1,0,2)
except in geometry $G_{13}$ and $G_{15}$ where the position of upper two detectors
and lower two detectors are interchanged from (3,1,0,2) to (1,3,2,0).
Geometries $G_{13}$ and $G_{14}$ are called Square geometries
since the distance between centers of top and bottom bars is about 95 cm, roughly equal to 
the length of the bars. The inter detector distances are kept almost equal.  
The physical picture of this geometry $G_{14}$, is given in Fig.~\ref{fig2det}.
Geometry $G_{11}$ corresponds to an increased distance of the top detector from bottom
detector to 110.7 cm is called Large geometry. In geometry $G_{17}$, we squeeze the setup
such that the distance between the top and bottom detectors is 63.6 cm keeping the
inter detector distances equal. This is referred to as Small geometry. 
Close geometries $G_{15}$ and $G_{16}$ are used to measure the efficiencies of the detectors
where the separation between two bars is kept very small as 3.2 cm. 
Table~\ref{table3} shows the vertical distances which are measured (from detector centers) 
from the bottom detector and calculated time taken by a light ray to traverse the vertical paths
from the top detector in the various geometries.

\begin{table}[ht]
\caption{The vertical distances (which are measured from center to center from the bottom detector)
 and calculated time taken by light in the setups (which are calculated from the top detector).}
\label{table3}
\begin{center}
\begin{tabular}{|c|c|c|c|c|c|} 
\hline
{\bf Geometry} & Distance/Time & {\bf Det$_3$} & {\bf Det$_1$} & {\bf Det$_0$} & {\bf Det$_2$} \\ 
\hline
$G_{14}$ (Square) & Distance (cm) & 95.0 & 63.3 & 31.8 & 0.0 \\
(3,1,0,2) & Time (ps) & 0.0 & 1056.6 & 2106.6 & 3166.6 \\
\hline
$G_{13}$ (Square) & Distance (cm) & 63.3 & 95.0 & 0.0 & 31.8\\
(1,3,2,0) & Time (ps) & 1056.6 & 0.0 & 3166.6 & 2106.6  \\
\hline
$G_{11}$ (Large) & Distance (cm) & 110.7 & 63.2 & 31.7 & 0.0 \\
(3,1,0,2) & Time (ps) & 0.0 & 1583.3 & 2633.3 & 3690.0  \\
\hline
$G_{17}$ (Small) & Distance (cm) & 63.6 & 42.4 & 19.45 & 0.0\\
(3,1,0,2) & Time (ps) & 0.0 & 706.7 & 1471.7 & 2120.0\\
\hline
\multicolumn{6}{|c|}{\bf Efficiency setups (Close setups)}   \\
\hline
$G_{15}$ (Close) &  Distance (cm)  & 18.5 & 27.8 & 0.0 & 9.2\\
(1,3,2,0) & Time (ps) & 310.0 & 0.0 & 926.7 & 620.0  \\
\hline
$G_{16}$ (Close) &  Distance (cm) & 27.8 & 18.5 & 9.2 & 0.0 \\
(3,1,0,2) & Time (ps) & 0.0 & 310.0 & 620.0 & 926.7\\
\hline
\multicolumn{6}{|c|}{\bf Position calibration setup}  \\
\hline
$G_{Cross}$ & Distance (cm) & 31.5 & 21.5 & 12.0 & 0.0\\
(3,1,0,2) & Time (ps) & 0.0 & 333.3 & 650.0 & 1050.0\\
\hline
\multicolumn{6}{|c|}{\bf Background setups}  \\
\multicolumn{6}{|c|}{Described in the result section}  \\
\hline
\end{tabular}
\end{center}
\end{table}

\section{Analysis methods}

\subsection{Time and position}

\begin{figure}
\centering
\includegraphics[width=0.95\textwidth]{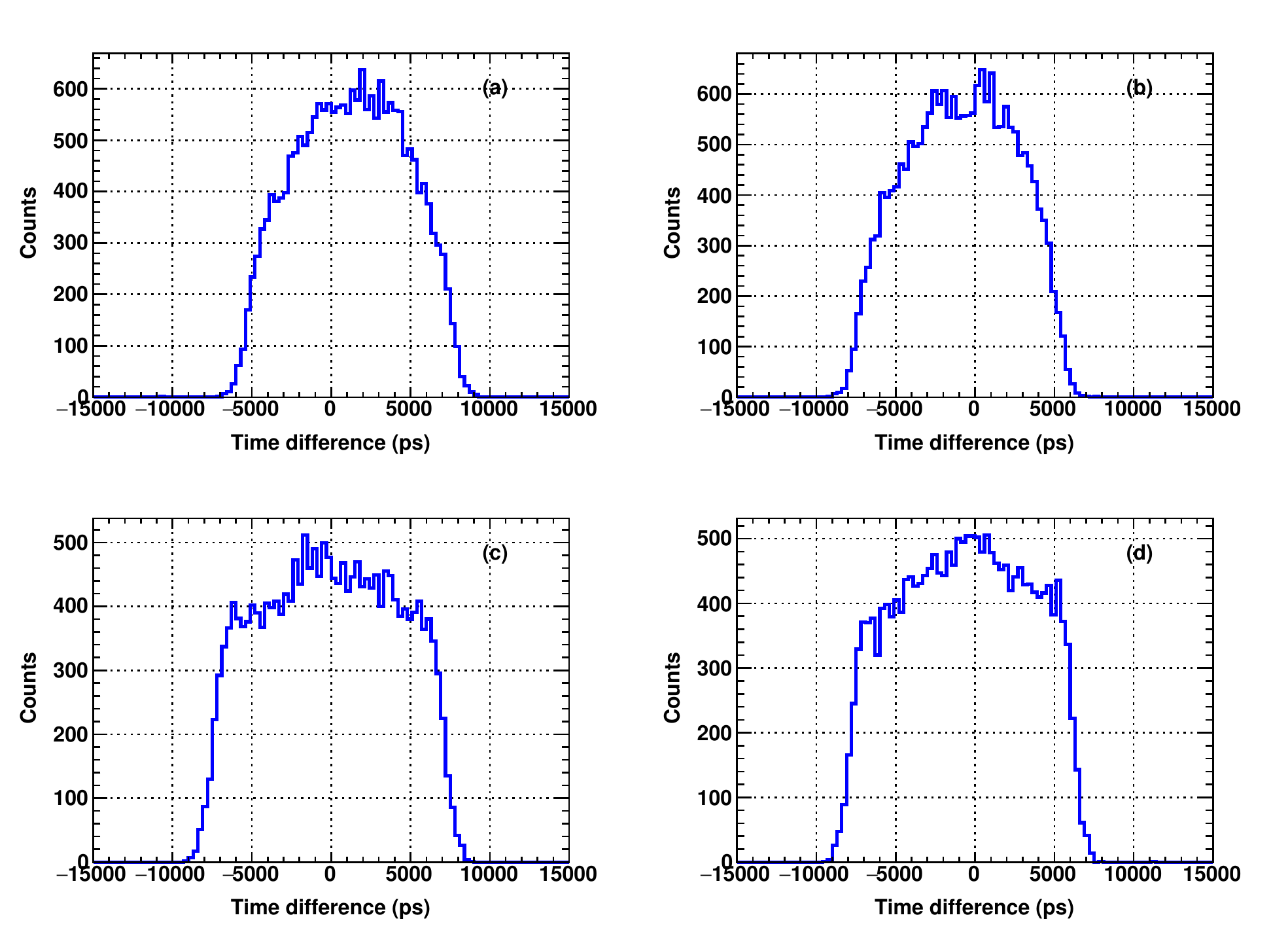}
\caption{Distribution of time difference between two ends of a detector, for each of the
  4 detectors for the case of muons in the Square geometry $G_{14}$. (a) Detector 0
  (b) Detector 1 (c) Detector 2 and (d) Detector 3. The events are under finalised cut 
  conditions.}
\label{Fig4TimeDetEnds}
\end{figure}

\begin{figure}
\centering
\includegraphics[width=12.5cm, height=2.5cm]{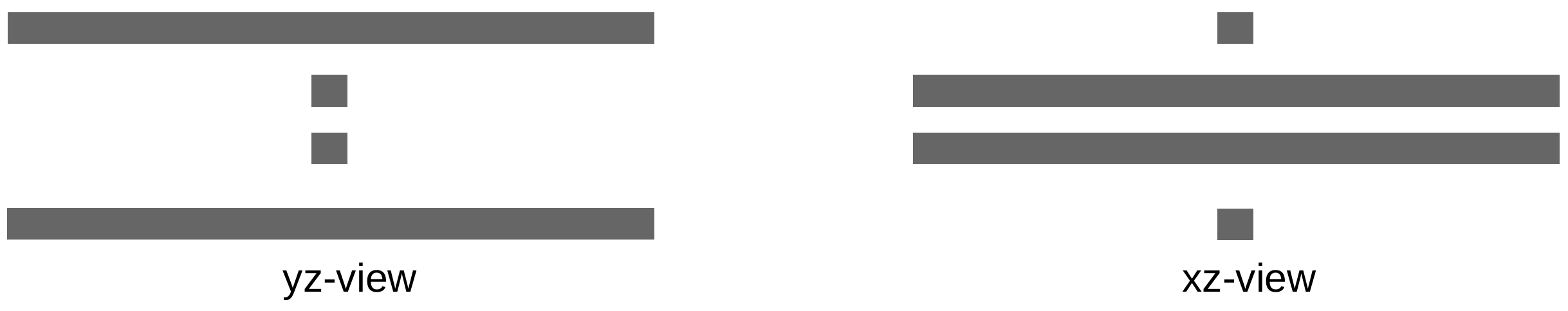}
\caption{Cross geometry $G_{Cross}$ of four detectors seen from two perpendicular sides
used to determine the time corresponding to muon positions
in the overlap region of detectors (5.6 cm $\times$ 5.6 cm).
The detectors are arranged in sequence (3,1,0,2) from top to bottom.
}
\label{Fig7CrossGeo}
\end{figure}

\begin{figure}
\centering
\includegraphics[width=0.85\textwidth]{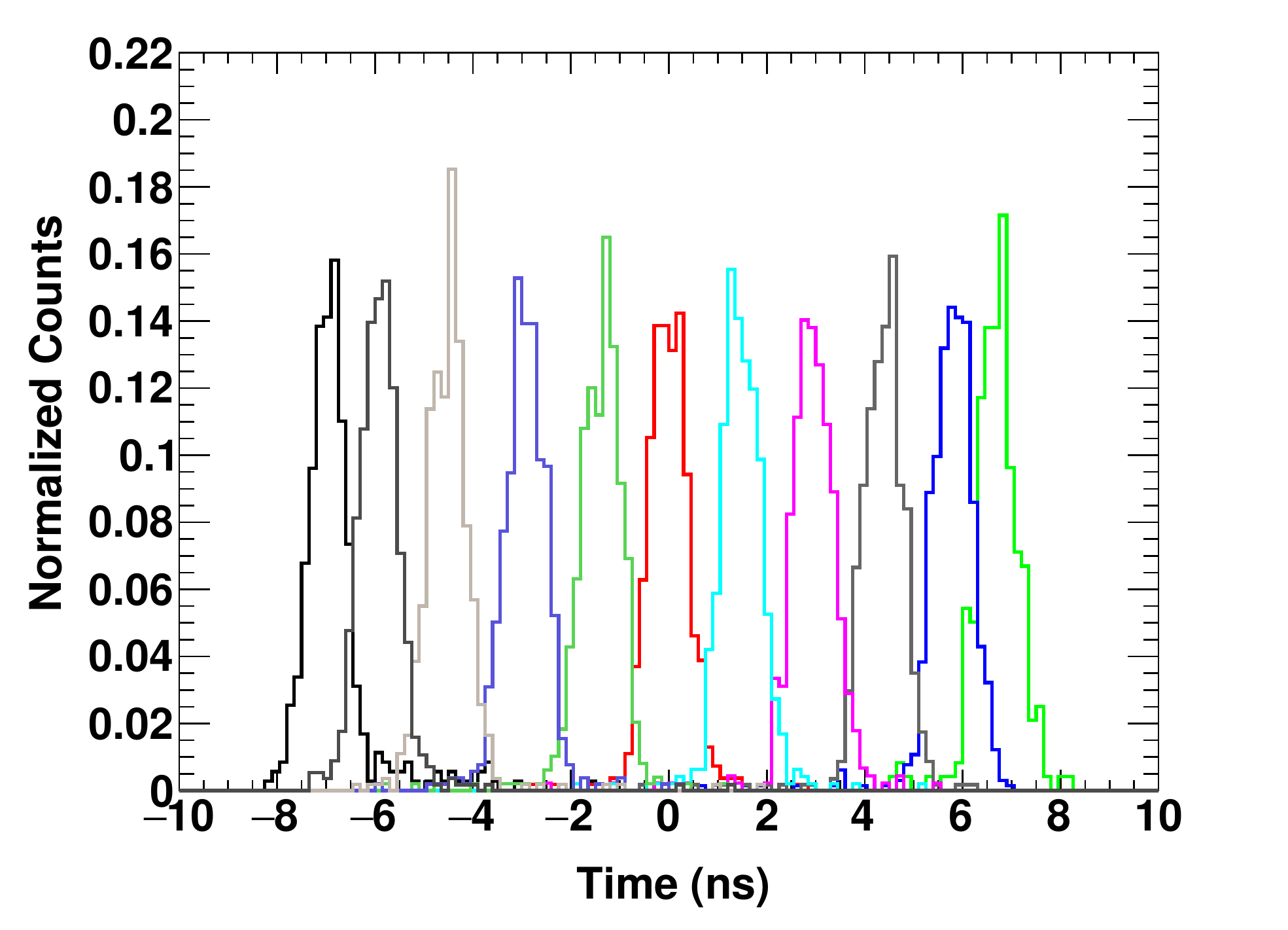}
\caption{The measured time for muon events at 11 different positions shown for one of the
  detectors. The positions in cm are  
  (-48.0, -40.1, -30.1, -20.1, -10.0, 0.0, 10.0, 20.1, 30.2, 40.3, 48.0).
}
\label{Fig6Time11Pos}
\end{figure}

\begin{figure*}
\begin{center}
\includegraphics[width=7.5cm, height=7.5cm]{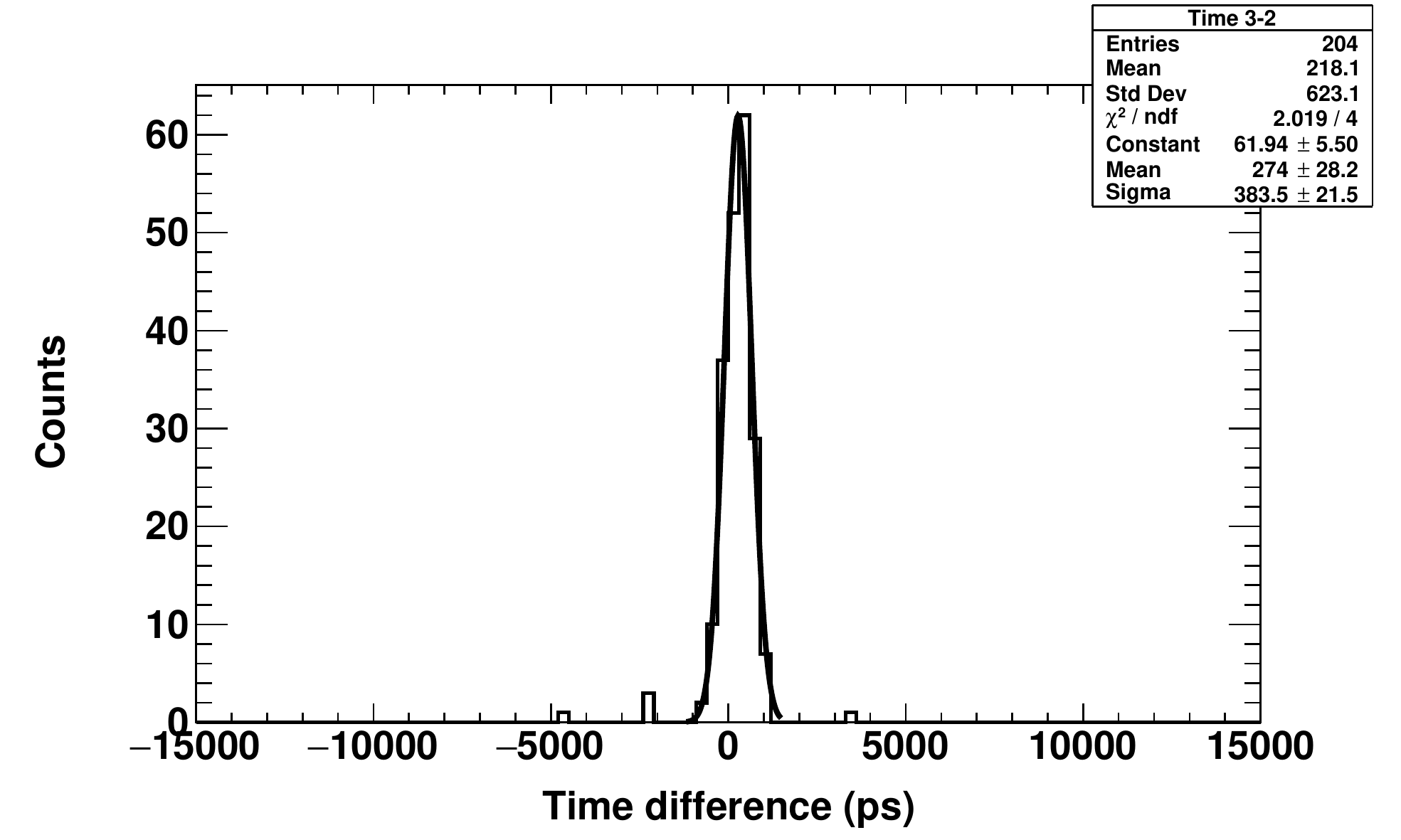}
\includegraphics[width=7.5cm, height=7.5cm]{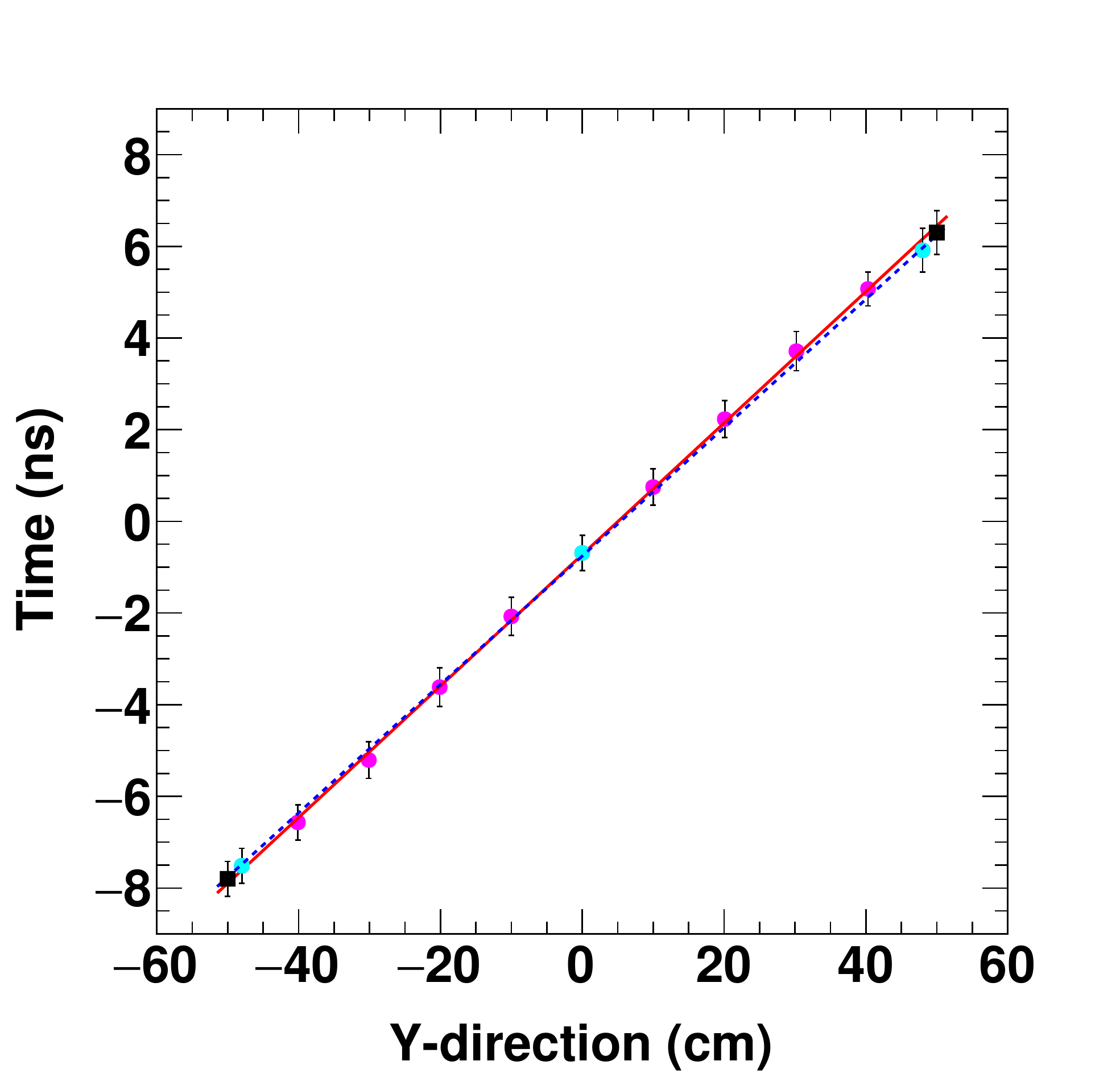}
\end{center}
\caption{(a) Left: Time resolution obtained using vertical muons at center of detector 1 using Cross geometry $G_{Cross}$.
  (b) Right: The measured time difference plotted as a function of 
 position of muon in scintillator bar at 11 positions.
 The solid line is a fit to 11 points and dashed line is a fit to three
 points (middle point and 2 last points) shown by blue.
 Two square black points at $\pm 50$ cm are the points where the time difference
 distributions between two ends of the detector bar fall to $50\%$ of their top value.
 }
\label{fig7TimeRes}
\end{figure*}

 The most important parameter in the measurements is the time difference between the
two ends of a detector which will finally be used for position measurement. 
In addition, the average time of the two sides of a detector is used as time arrival of the signal
in that detector and will be used for inter detector time difference which is very
useful to define a time window of an event and to know the directionality of track. 
Figure~\ref{Fig4TimeDetEnds} shows the distributions of time difference between two
ends of a detector, for each of the 4 detectors for the case of muons in the
Square geometry $G_{14}$ for (a) Detector 0 (b) Detector 1 (c) Detector 2 and (d) Detector 3.
The events are filled under the coincidence in all 4 detectors within 7 ns time
from top to bottom detectors under finalised cut conditions which are explained
in next section and we produce the values here. The minimum energy deposition
cut Ecut=600 ChNo. (8.07 MeV) and the conditions on 
track quality parameter used are $\chi^{2}$/NDF<3, 
$\Delta t_{rms}$<0.9, $E_{rms}$<0.2 and PSD<0.78.
Time difference distributions are concentrated more towards centers in detector 0 
and 1 as they are kept in the middle of the setup.

 We use vertical muons to get the position time relation using 
two bars kept in cross position to the other two bars all in coincidence and then moving
two of them to change the position.
Figure~\ref{Fig7CrossGeo} shows the Cross geometry $G_{Cross}$ of four detectors seen from
two perpendicular sides which are used to determine the time corresponding to muon positions
in the overlap region of detectors (5.6 cm $\times$ 5.6 cm).
The detectors are arranged in sequence (3,1,0,2) from top to bottom.

Figure~\ref{Fig6Time11Pos} shows the measured time for muon events
at 11 different positions (center of overlap area) of the detector given in cm as 
(-48.0, -40.1, -30.1, -20.1, -10.0, 0.0, 10.0, 20.1, 30.2, 40.3, 48.0).
Here 11 histograms are plotted on a single canvas.
Figure~\ref{fig7TimeRes}(a) shows the measured time resolution $\sigma_t$ using vertical
muons which is 383 ps. This corresponds to a position resolution of $\approx$ 3 cm from
a single bar which is reduced to 1.5 cm when we use four bars for tracking. 
Figure~\ref{fig7TimeRes}(b) shows the measured time difference plotted as a function of
position of muon in scintillator bars at 11 positions.
The solid line is a fit to 11 points
obtained as $t ({\rm in\,\,ns}) = 0.143\times y ({\rm in\,\,cm}) -0.725$.
The dashed line is a fit to three points
(middle point and 2 last points) shown by blue circle which is
obtained as $t ({\rm in\,\,ns}) = 0.140\times y ({\rm in\,\,cm}) -0.757$.
 The two lines give two calibrations for position and time which are
then included in systematics uncertainties in flux measurements.
 Two black points at $\pm 50$ cm are the points where the time difference
distributions between two ends of a detector (of Figure~\ref{Fig4TimeDetEnds}) fall to $50\%$ of their top value.
These points fall on the fitted lines and hence
can also be used as a quick alternative method to obtain time-position calibration.

Table~\ref{table4} first row shows the time correction for the center of each detector.
Table~\ref{table4} second row shows inter detector time offset with respect to top detector (Det$_3$).
The inter detector time offset is the time duration of the events between two
detectors is also measured using vertical muons assuming them traveling with
speed of light.
The events are in coincidence when all the 8 signals from the four bars
arrive within 25 ns. After all the corrections are done we are in a position
to tighten the coincidence condition to have a time difference within 7 ns 
(5 ns in case of Close geometries $G_{15}$ and $G_{16}$) in the top and bottom detector.

\begin{table}[ht]
\caption{Time correction for the center of each detector. The second line 
shows the inter detector timing offset with respect to top detector (Det$_3$).}
\label{table4}
\begin{center}
\begin{tabular}{|c|c|c|c|c|} 
\hline
{\bf Time (ps)} & {\bf Det$_0$} & {\bf Det$_1$} & {\bf Det$_2$} & {\bf Det$_3$} \\ 
\hline
$T_{\rm center}$ & 1194.0 & -931.0 & -37.0 & -689.0\\
\hline
\hline
$T_{\rm offset}$ & 1167.3 & 681.8 & -340.4 & 0.0 \\
\hline
\end{tabular}
\end{center}
\end{table}

\subsection{Energy deposition}

\begin{figure}
\centering
\includegraphics[width=0.60\textwidth, height=0.5\textwidth]{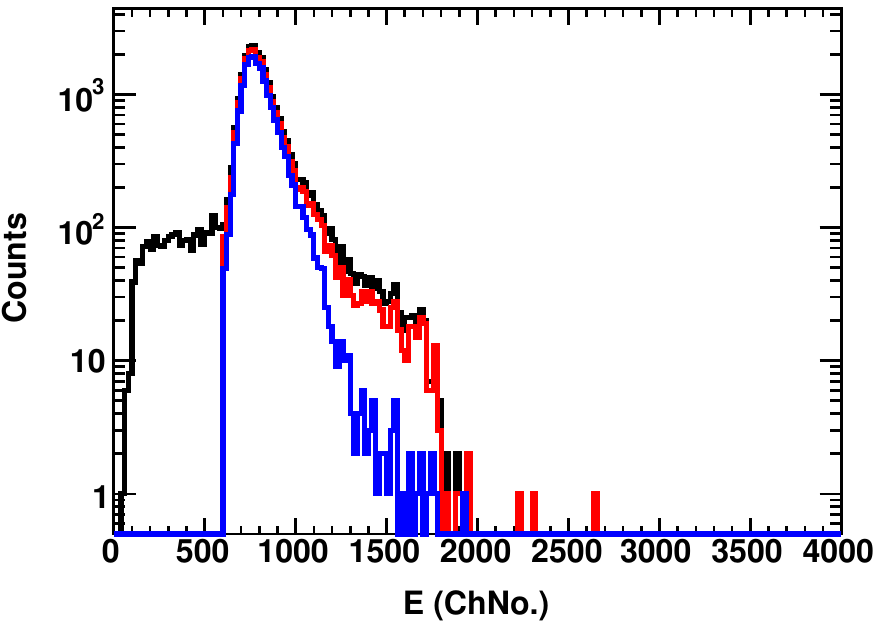}
\caption{Geometric mean of the charge collection (in Channel number ChNo.) at the two ends of one of the 
  detector in Square geometry $G_{14}$ for Detector 2
  (i) Black histogram corresponds to the coincidence of events within 7 ns with energy deposition
  $>$ 0 ChNo. in all 4 detectors. 
  (ii) Red histogram is the coincidence of events within 7 ns and with minimum
  energy deposition $E>600$ ChNo. in all 4 detectors and
  (iii) Blue histogram is the coincidence of events within 7 ns and with minimum energy deposition
  $E>600$ ChNo. in all 4 detectors and after cleaning muon events using track
  quality cuts defined later in the next subsections. 
}
\label{Fig8EGMa}
\end{figure}

 To construct the energy deposition $E$ of the signal we take geometric mean of the two integrated
charges $E_1$ and $E_2$ obtained at the two ends of the bars as $E=\sqrt{E_1\,E_2}$.
Since there is attenuation of the signal as it travels in the scintillator, the charge at
the end where the event is closer will be more than that at the other end. Geometric mean
makes the signal position independent. Signal gains are matched offline in all four detectors.
Figure~\ref{Fig8EGMa} shows the geometric mean 
of the charge collection at the two ends of one of the detectors in the Square geometry $G_{14}$ for
Detector 2.
 Here, (i) Black histogram corresponds to the coincidence of events within 7 ns with energy deposition $>$0 ChNo. in all 4 detectors. 
  (ii) Red histogram is the coincidence of events within 7 ns and with minimum
  energy deposition $E>600$ ChNo. in all 4 detectors and
  (iii) Blue histogram is the coincidence of events within 7 ns and with minimum energy deposition
  $E>600$ ChNo. in all 4 detectors and after cleaning muon events using track
  quality cuts defined later in the next subsections. 

In the analysis, $E>600$ ChNo. cut is placed on energy deposition where we see a step in the
black histogram showing the starting of muon peak. The value of this cut is obtained 
such that very small number of muon events are lost, the method is presented in the next section.
The energy calibration to convert channel number (ChNo.) with the energy in MeV 
is done after the tracking schemes are finalized in the next subsections.  

 Energy deposition is measured using Cross geometry $G_{Cross}$ (for vertical muons) of detectors.
The energy deposition by muons in each detector is fitted using the
Langau (Landau+Gaussian) function. The ratios of Most Probable Value (MPV) for a
detector and detector 0 is the energy gains of that detector with respect to
detector 0.    
 The energy deposition gain matching with respect to detector 0 are
1.11, 1.10 and 1.04 for detectors 1, 2 and 3 respectively.
Table~\ref{table5} shows the parameters of Langau (Landu+Gaussian) distribution
function which was obtained by fitting energy deposition distribution with 
all final cuts in the detector 0 by vertical muons measured using Cross geometry $G_{Cross}$.
This distribution of deposited energy is used as input in the
simulation and these parameters are also used in the energy calibration.

\begin{table}[ht]
\caption{Langau parameters obtained by fitting energy deposition in the 
detector 0 by vertical muons measured using Cross geometry $G_{Cross}$.}
\label{table5}
\begin{center}
\begin{tabular}{|c|c|c|c|c|} 
\hline
{\bf Detector} & {\bf Width} & {\bf MPV} & {\bf Area} & {\bf GSigma}\\ 
\hline
0 & 31.9 & 746.8 & 9256.3 & 32.2\\ 
\hline
\end{tabular}
\end{center}
\end{table}

\subsection{Pulse Shape Discrimination (PSD)}
Liquid scintillator has different pulse shapes of energy 
deposition for different particles which gives a unique handle of 
Pulse Shape Discrimination (PSD) as compared to other detectors.
PSD parameter is defined as the ratio of difference of
energy depositions by a particle in the detector
in the Long gate (200 ns) and Short gate to the energy deposition by
particle in Long gate as
\begin{equation} \label{eq0}
 {\rm PSD} = \frac{E_{\rm Long} - E_{\rm Short}}{E_{\rm Long}}.
\end{equation}

\begin{figure*}
\begin{center}
\includegraphics[width=7.5cm, height=7.5cm]{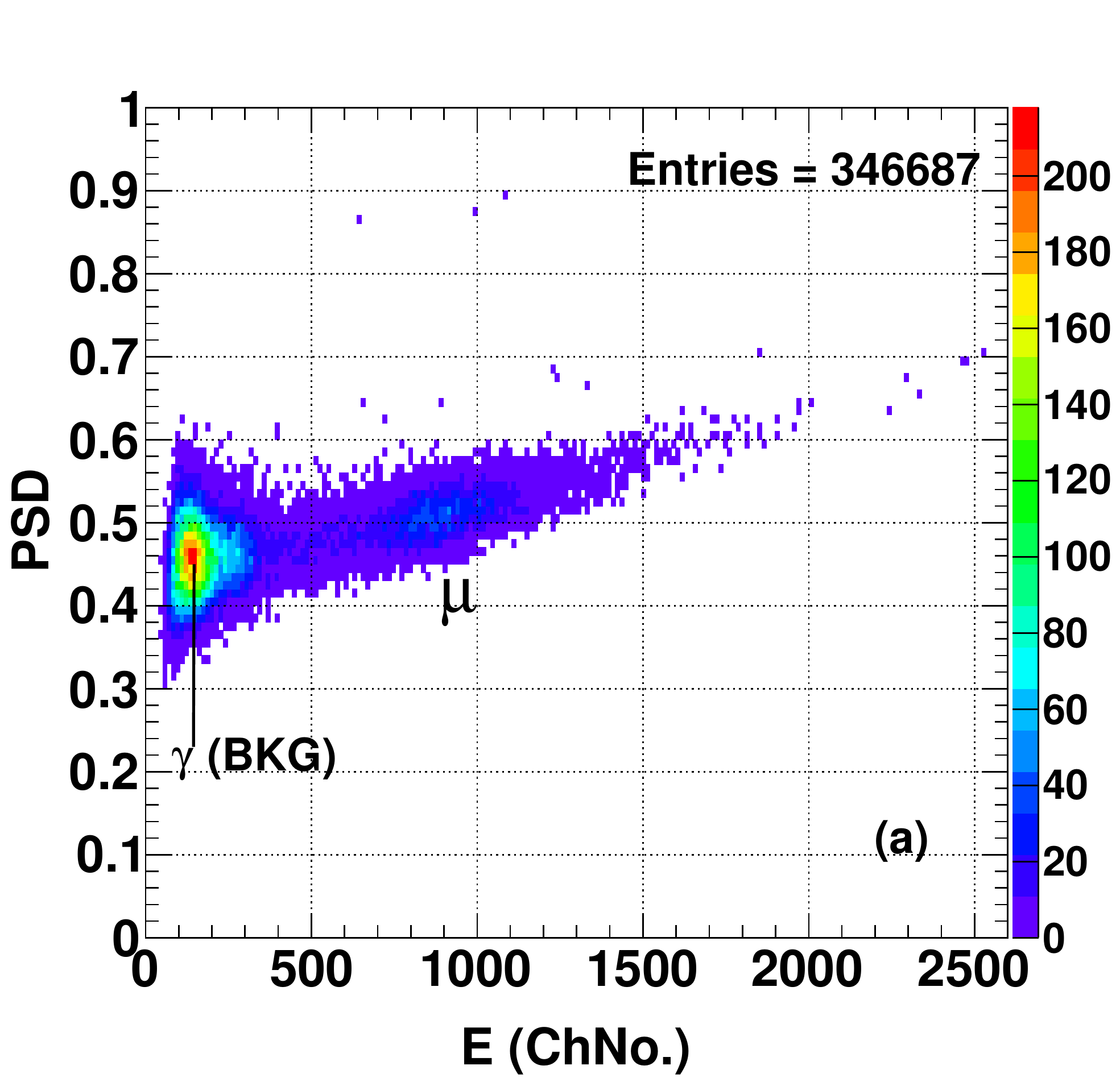}
\includegraphics[width=7.5cm, height=7.5cm]{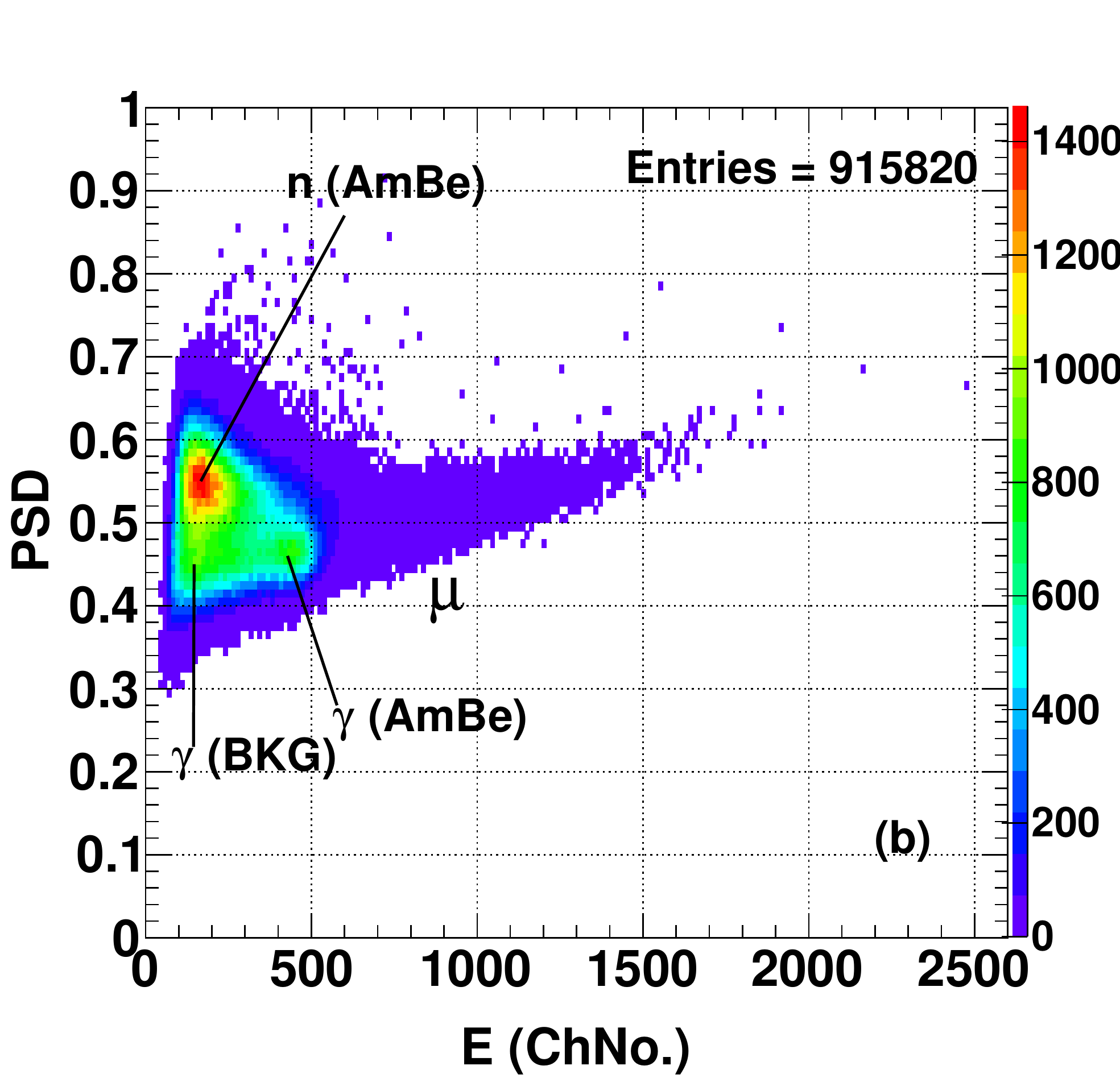}
\end{center}
\caption{2D histograms of PSD versus energy for Detector 1 which is run in 
singles mode (a) Without AmBe source showing background particles 
and muon bump (b) With AmBe source showing neutrons and photons 
imposed over background and muons. Both the spectra are built 
over same acquisition time of $\sim$13 minutes.}
\label{Fig9EdepVsPSD}
\end{figure*}

Figure~\ref{Fig9EdepVsPSD} shows 2D histograms of PSD versus energy 
for Detector 1 which is run in singles mode  (a) Without AmBe source showing 
background particles and muon bump  (b) With AmBe source showing 
neutrons and photons imposed over background and muons. Both the 
spectra are built over same acquisition time of $\sim$13 minutes. 
One can clearly see that the neutrons have higher PSD as compared 
to photons. A lower threshold on  energy cut on all 4 scintillators
gives a clean muon spectrum. We also apply a PSD cut which removes 
heavy particles if any. The finalized PSD cuts obtained from this 
figure are PSD $<0.6$ for $G_{15}$, $G_{16}$ and $G_{17}$ where 
the Short gate value is 24 ns. Similary PSD $<0.78$ is determined
for all other geometries where the Short gate value is 20 ns.

\subsection{Track quality parameters}

\begin{figure}
\centering
\includegraphics[width=0.85\textwidth]{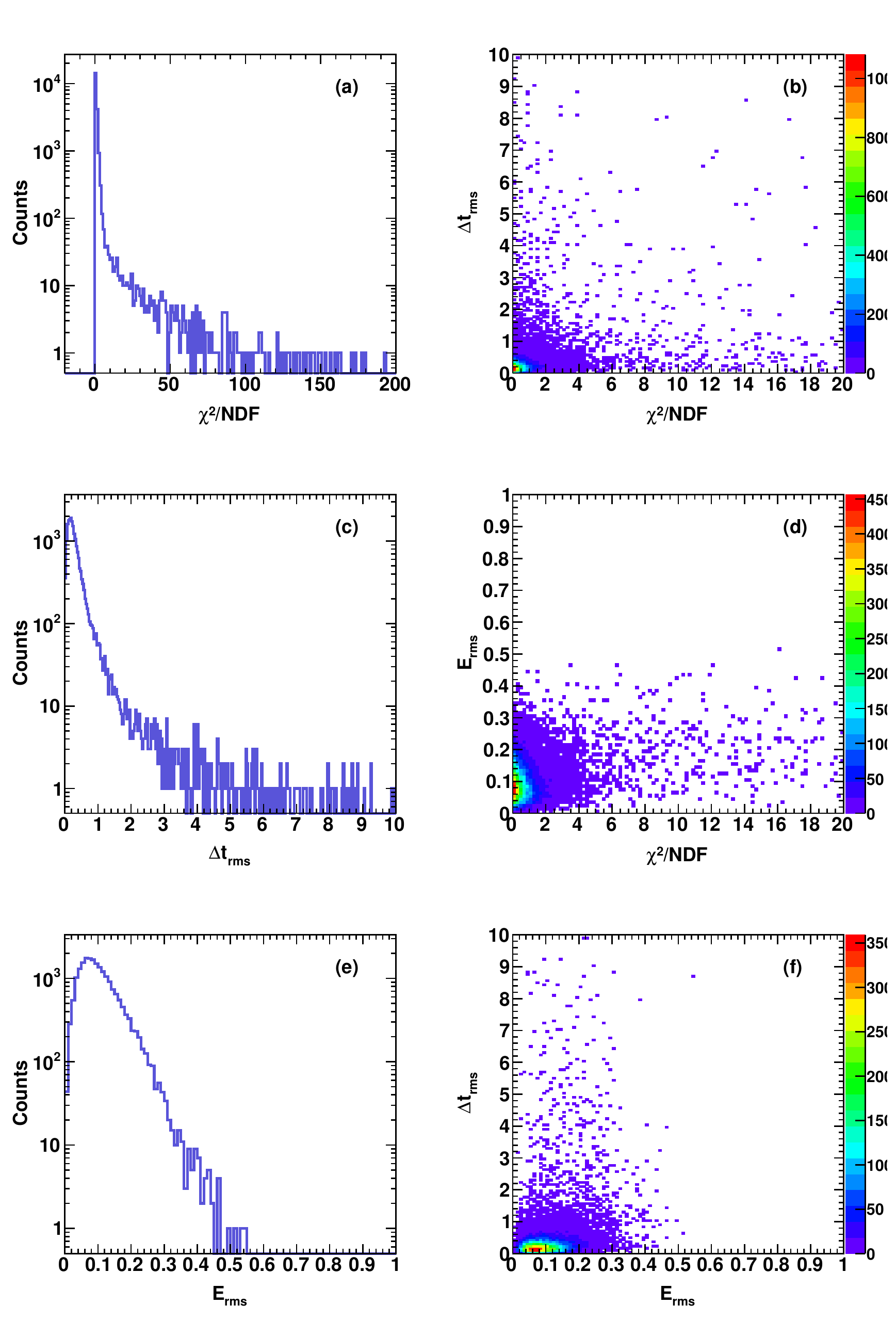}
\caption{Distributions of track quality parameters
and their correlations, for all the coincident events (within 7 ns) in the detectors
with finalised cut on energy deposition and PSD in the Square geometry $G_{14}$,
(a) $\chi^{2}$/NDF (b) $\chi^{2}$/NDF with $\Delta t_{rms}$ (c) $\Delta t_{rms}$ 
(d) $\chi^{2}$/NDF with $E_{rms}$ (e) $E_{rms}$ and (f) $\Delta t_{rms}$ with $E_{rms}$.
It is clear from the correlation histograms (b), (d) and (f) that the track
quality parameters are independent from each other and background can be reduced 
by applying cuts on each of the parameters.}
\label{Fig10TrackDist}
\end{figure}

In this analysis, we devise three dimensionless track quality parameters to reject the background
events in $m$ detectors. One is the usual $\chi^{2}$/NDF. The other two are devised such that
they favour tracks that take similar time to travel equal inter detector distances and
deposit similar energy in all detectors.
\begin{enumerate}

\item $\chi^{2}$/NDF: This is calculated by fitting a straight
line to the $m$ measured data points in $m$ detectors.

\item $\Delta t_{rms}$: If $t_{0}$, $t_{1}$, $t_{2}$ and $t_{3}$ are the timings
of events in the detectors starting from top to bottom then

\begin{eqnarray} \label{eq1}
   \Delta t_{rms} &= & \sqrt{ \frac{1}{m-1} \sum_{i=1}^{m-1} \Bigg(\frac{\Delta t_{i} - \overline{\Delta t}}{\Delta t_{i}}
    \Bigg)^{2}}, \\
 \,\, {\rm where} \,\,\,  \Delta t_{i} & =  &\frac{t_{i}-t_{i-1}}{(t_{i}-t_{i-1})_{\rm expected}} \\
 \,\, {\rm and} \,\,  \overline{\Delta t} &  = & \frac{1}{m-1} \sum_{i=1}^{m-1} \Delta t_{i}.
\end{eqnarray}
Here $(t_{i}-t_{i-1})_{\rm expected}$ is the time taken by the light from $(i-1)^{th}$ to $i^{th}$ detector in downward sequence.

\item $E_{rms}$:
If $E_{0}$, $E_{1}$, $E_{2}$ and $E_{3}$ are the energy depositions of events in the detectors
0, 1, 2 and 3 respectively then 

\begin{eqnarray}
 E_{rms} &= & \sqrt{ \frac{1}{m} \sum_{i=0}^{m-1} \Bigg(\frac{E_{i} - \overline{E}}{E_{i}}  \Bigg)^{2}  }, \\
{\rm where} \,\,\,
  \overline{E} & = &\frac{1}{m} \sum_{i=0}^{m-1} E_{i}.
\end{eqnarray}

\end{enumerate}

Figure~\ref{Fig10TrackDist} shows the distributions of track quality parameters
and their correlations, for all the coincident events (within 7 ns) in the detectors
with finalised cut on energy deposition and PSD in the Square geometry $G_{14}$,
(a) $\chi^{2}$/NDF (b) $\chi^{2}$/NDF with $\Delta t_{rms}$ (c) $\Delta t_{rms}$ 
(d) $\chi^{2}$/NDF with $E_{rms}$ (e) $E_{rms}$ and (f) $\Delta t_{rms}$ with $E_{rms}$.
It is clear from the correlation histograms (b), (d) and (f) that the track
quality parameters are independent from each other and background can be reduced 
by applying cuts on each of the parameters.

\subsection{Energy calibration}

\begin{figure}
\centering
\includegraphics[width=0.98\textwidth]{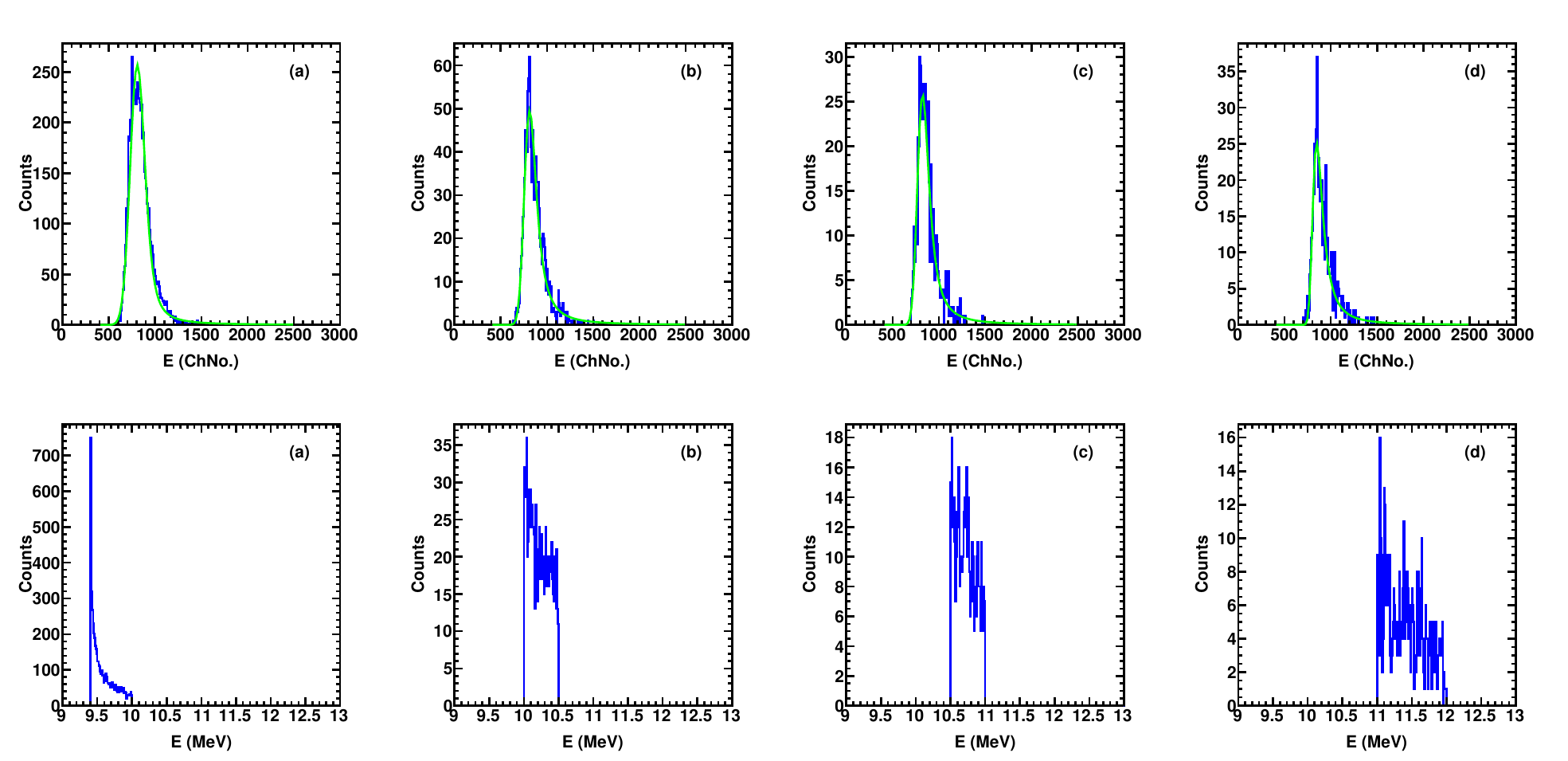}
\caption{The energy spectra in channel number (top panels) in different
bins of energy (lower panels) calculated using measured pathlengths traversed by muons
in the detector. The energy bins used are 9.4-10 MeV,
10-10.5 MeV, 10.5-11 MeV and 11-12 MeV in the Square geometry $G_{14}$.
Events are under finalised cut conditions as mentioned in the next section.
}
\label{Fig11ECal}
\end{figure}

\begin{figure}
\centering
\includegraphics[width=8.0cm, height=8.0cm]{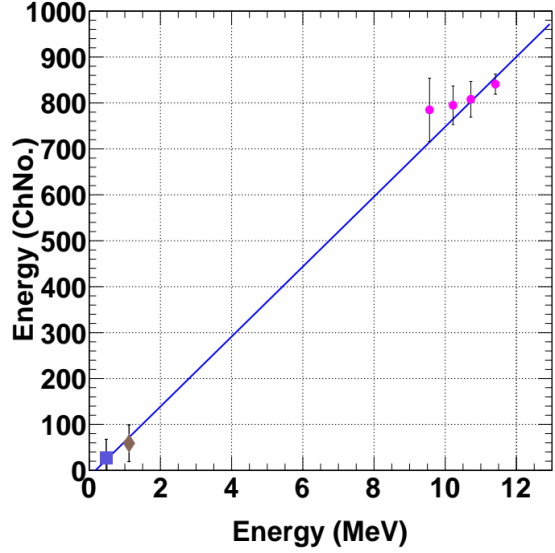}
\caption{The peak channel number (MPV) of the energy spectra
  plotted as a function of mean energy in MeV calculated for each bin
  obtained using measured pathlength in the detector corresponding to
  tracks with different angles. The two lowest points are from the Compton
  edge of $\gamma$-ray from $^{137}$Cs source (0.662 MeV) and $^{60}$Co (1.332 MeV). 
 }
\label{Fig12ChMeV}
\end{figure}

 The detector width (active material) is 4.9 cm and thus energy loss expected for a vertical muon
in the detector bar is $E_{\rm Vertical}=9.4$ MeV considering an energy loss of 1.92 MeV/cm for
a muon of energy 2 GeV. 
 Since we can reconstruct angle ($\theta$) of the track and hence path lengths in a detector and
thus we can calculate the energy deposits in MeV for a track at angle $\theta$ as
\begin{equation}
  E(\theta) = \frac{E_{\rm Vertical}}{\cos\theta}
\label{enerpath}
\end{equation}  
 This allows us to fill energy spectrum in channel numbers in the required bins of energy
in MeV as per Eq~(\ref{enerpath}) constrained using pathlengths (measured using zenith angle of track). 
The channel number then can be calibrated in terms of energy. 
Figure~\ref{Fig11ECal} shows the energy spectra in channel number (top panels) in different
bins of energy (lower panels) obtained using measured pathlengths traversed by muons
in the detector. The events are under finalised cut conditions as mentioned in the next section.
The energy bins used are 9.4-10 MeV, 10-10.5 MeV, 10.5-11 MeV, 11-12 MeV.
 The energy spectra in each bin are fitted with a Langau function and MPV is obtained.
 Figure~\ref{Fig12ChMeV} shows the peak channel number (MPV) of the energy spectra
  plotted as a function of mean energy in MeV calculated for each bin
  obtained using measured pathlength in the detector corresponding to
  tracks with different angles. The two lowest points are from the Compton
  edge of $\gamma$-ray from $^{137}$Cs source (0.662 MeV) and $^{60}$Co (1.332 MeV). 
The straight line is fitted including all points is given by 
\begin{equation}
 E_{\rm Ch} = m \times E_{\rm MeV} + c, 
\end{equation}
where $m=76$ ChNo./MeV and $c=-13$ ChNo.
Now we can calculate the energy cut in our analysis in MeV.
Thus 600 ChNo., 720 ChNo. and 1600 ChNo. are equal to 8.07 MeV, 9.64 MeV and 21.22 MeV respectively.

\section{Measurement of efficiencies}

 We obtain the efficiency of one of the middle detectors by using coincidence with
the other 3 detectors with stringent cuts on energy deposition, PSD and track quality cuts.
In the energy deposition range $E_{\rm cut}<E<$1600 ChNo. (including PSD),
the efficiency of $E_{\rm cut}$ of detector 1
in Close geometry $G_{16}$ is calculated as
\begin{equation}
\epsilon_{E1} = \frac{\rm Counts\,\,in\,\, (0\times1\times2\times3)}{\rm Counts\,\,in\,\, (0\times2\times3)} 
\label{effeq}
\end{equation}  
where stringent cuts are applied on other 3 detectors (0,2,3) to have incident
muons which are background free.
The values of stringent cuts on energy deposition, PSD and Track quality cuts are 
720 ChNo.$<E<$1600 ChNo., PSD$<$0.6 ($<$0.78 for Square and Large geometries),
$\chi^{2}$/NDF < 0.3, $\Delta t$ < 0.3 and $E_{rms}$ < 0.12 respectively.
This way efficiencies of two detectors 0 and 1 can be obtained in Close geometry 
$G_{16}$ where they are placed in the middle.
The geometry $G_{15}$ (reverse of $G_{16}$) has been used to obtain
efficiencies of detectors 2 and 3. In both the geometries
$G_{15}$ and $G_{16}$, the detectors were placed with a nominal separation
of 3.2 cm among them to have an excellent alignment of all four bars.
 Later we also measure the efficiencies of middle detectors in Square geometries $G_{13}$
and $G_{14}$. The efficiencies come down which now include the misalignment among the 
four detectors. The efficiencies of top and bottom detectors are taken from Close
geometries $G_{15}$ and $G_{16}$ but the efficiencies of middle detectors are always
taken from actual geometry of measurement. This includes the effect of possible
misalignment for geometries where the inter detector distances are large.

\begin{figure}
\centering
\includegraphics[width=1.0\textwidth]{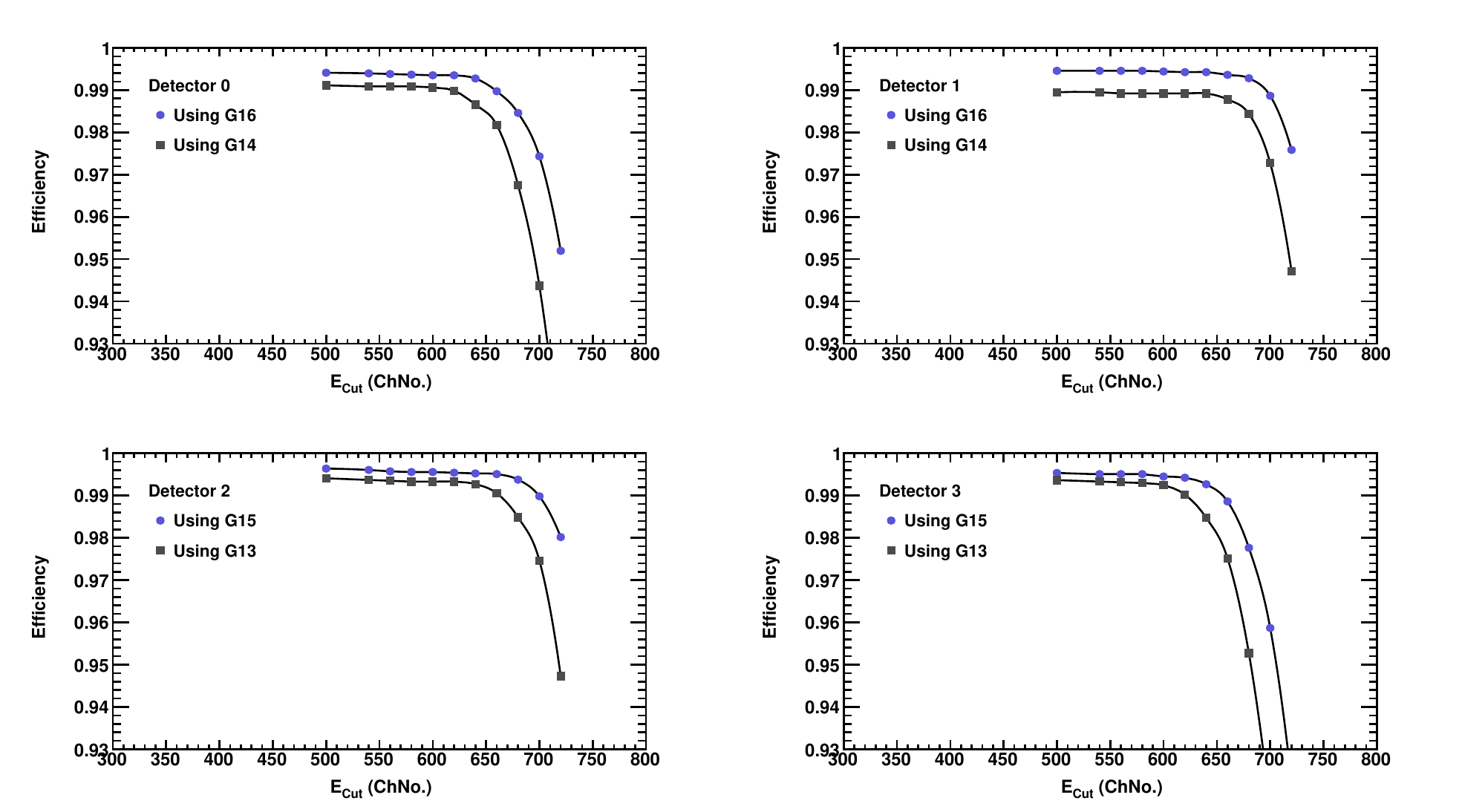}
\caption{Efficiencies of detectors as a function of energy deposition for all
4 detectors using Close geometry $G_{15}$ and $G_{16}$ and Square geometries $G_{13}$ 
and $G_{14}$. Here the cut is applied as $E_{Cut}<E<1600$ ChNo.}
\label{Fig13EnEff}
\end{figure}
 
Figure~\ref{Fig13EnEff} shows the efficiencies of detectors as a function of
energy deposition cuts $E_{\rm cut}$ in $E_{\rm cut}<E<$1600 ChNo.. 
Efficiencies of detectors for various energy
deposition cuts were obtained using geometries $G_{15}$, $G_{16}$, $G_{13}$ and $G_{14}$.
With the help of this graph, we have chosen the finalised value of energy 
cut $E_{\rm cut}$ equal to 600 ChNo. (8.07 MeV)
to reduce the background by signal ratio without compromising much on genuine
muon events.

To measure the efficiency of track quality parameters we use a data driven method. Here
to get the efficiency of the cut on a parameter under study, the other 
parameters and energy deposition cuts were kept stringent. 

\begin{figure}
\centering
\includegraphics[width=1.0\textwidth]{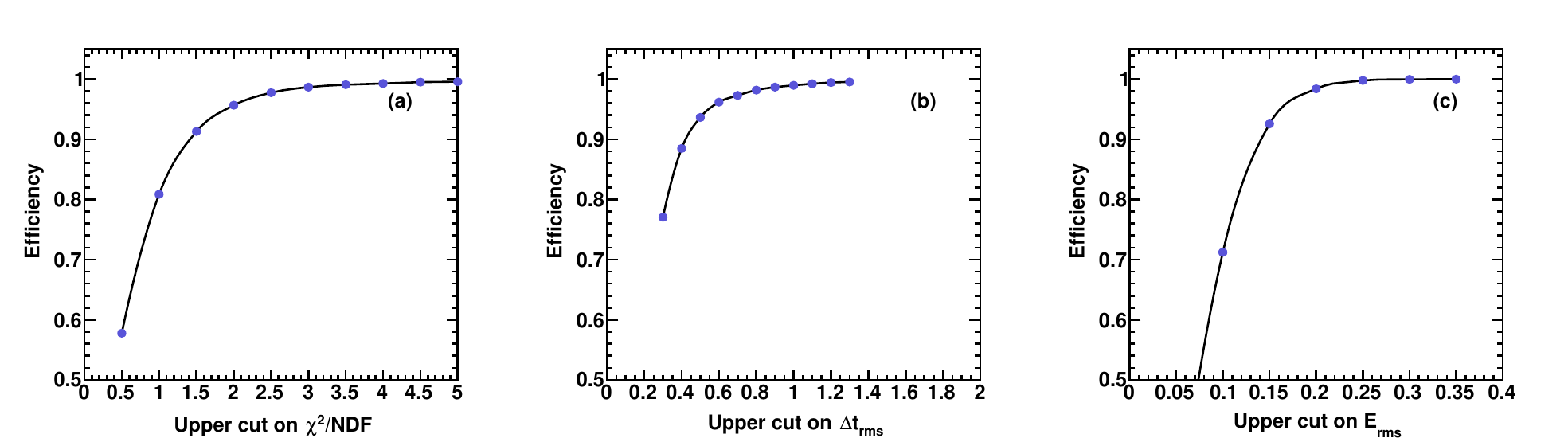}
\caption{Efficiencies of track quality parameters as a function of 
the cuts for Square geometry $G_{14}$ (a) Upper cut 
on $\chi^{2}$/NDF (b) Upper cut on $\Delta t_{rms}$ and (c) Upper cut on $E_{rms}$.}
\label{Fig14EffTrack}
\end{figure}

Figure~\ref{Fig14EffTrack} shows the efficiencies of track quality
parameters as a function of the cuts for Square geometry $G_{14}$
(a) Upper cut on $\chi^{2}$/NDF (b) Upper cut on $\Delta t_{rms}$ and (c) Upper cut on $E_{rms}$.
With the help of these graphs we choose the values of tightest cuts but still be
close to plateau region in order to keep the efficiency high. 
We have chosen the finalised values of tracking cuts $\chi^{2}$/NDF$<$3, 
$\Delta t_{rms}<$0.9 and $E_{rms}<$0.2 for Square geometry $G_{14}$. For geometries 
$G_{13}$, $G_{14-2}$, $G_{14-3}$ and $G_{11}$, the values of tracking cuts are
kept same as those for $G_{14}$.
The finalised values of tracking cuts for Small geometry $G_{17}$ are chosen as
$\chi^{2}$/NDF$<$3, $\Delta t_{rms}<$2.5 and $E_{rms}<$0.2. The finalised 
value of tracking cuts for Close geometries ($G_{15}$ and $G_{16}$) are chosen as 
$\chi^{2}$/NDF$<$3, $\Delta t_{rms}<$2.5 and $E_{rms}<$0.22. In Small and Close
geometries it was required to loosen the $\Delta t_{rms}$ cut because of
limited time resolution of the detectors.

\begin{table}[ht]
\caption{Efficiency of detectors in various geometries for muon detection at finalised
energy deposition above $E_{\rm cut}$ = 8.07 MeV (including PSD cut).}
\label{table6}
\begin{center}
\resizebox{\columnwidth}{!}{%
\begin{tabular}{|c|c|c|c|c|} 
\hline
{\bf Geometry} & {\bf Det$_0$} & {\bf Det$_1$} & {\bf Det$_2$} & {\bf Det$_3$}\\ 
\hline
$G_{14}$ & 0.991 $\pm$ 0.001 & 0.989 $\pm$ 0.002 & 0.996 $\pm$ 0.001 & 0.994 $\pm$ 0.001\\
\hline
$G_{13}$ & 0.993 $\pm$ 0.001 & 0.994 $\pm$ 0.001 & 0.993 $\pm$ 0.001 & 0.992 $\pm$ 0.001\\
\hline
$G_{15}$ \& $G_{16}$ & 0.993 $\pm$ 0.001 & 0.994 $\pm$ 0.001 & 0.996 $\pm$ 0.001 & 0.994 $\pm$ 0.001\\
\hline
\end{tabular}
}
\end{center}
\end{table}

Table~\ref{table6} shows the efficiency of detectors in various geometries for muon detection at 
finalised energy deposition above $E_{\rm cut}$ = 8.07 MeV (including PSD cut) using Eq~\ref{effeq}.

\begin{table}[ht]
  \caption{Efficiency of finalised tracking cuts for various geometries.}
\label{table7}
  
\begin{center}
\begin{tabular}{|c|c|c|c|} 
  \hline
  {\bf Geometry}  & {$\bf \epsilon_1$}  & {$\bf \epsilon_2$}   &  {$\bf \epsilon_3$} \\ 
   & ({$\chi^{2}/NDF$}) & ({$\Delta t_{rms}$}) & ({$E_{rms}$})\\

\hline
$G_{13}$ & 0.986 $\pm$ 0.001 & 0.989 $\pm$ 0.001 & 0.987 $\pm$ 0.002 \\
\hline
$G_{14}$ & 0.987 $\pm$ 0.001 & 0.987 $\pm$ 0.002 & 0.984 $\pm$ 0.002 \\
\hline
$G_{11}$ & 0.985 $\pm$ 0.002 & 0.986 $\pm$ 0.002 & 0.986 $\pm$ 0.002 \\
\hline
$G_{17}$ & 0.992 $\pm$ 0.001 & 0.971 $\pm$ 0.002 & 0.986 $\pm$ 0.002 \\
\hline
$G_{14-2}$ & 0.971 $\pm$ 0.002 & 0.994 $\pm$ 0.001 & 0.980 $\pm$ 0.002 \\
\hline
$G_{14-3}$ & 0.974 $\pm$ 0.002 & 0.988 $\pm$ 0.002 & 0.977 $\pm$ 0.002 \\
\hline
$G_{15}$ & 0.984 $\pm$ 0.001 & 0.992 $\pm$ 0.001 & 0.987 $\pm$ 0.002 \\
\hline
$G_{16}$ & 0.987 $\pm$ 0.001 & 0.983 $\pm$ 0.001 & 0.984 $\pm$ 0.002 \\
\hline
\end{tabular}
\end{center}
\end{table}

Table~\ref{table7} shows efficiency of finalised tracking cuts for various geometries.
Efficiency errors in Tables~\ref{table6} and \ref{table7} are calculated using 
binomial error $\sigma_{\epsilon} = \sqrt{\epsilon (1- \epsilon)/N}$,
where $N$ is the total number of incident events and $\epsilon$ is the efficiency
of the optimised cut. 

\section{Monte Carlo simulation and geometrical acceptance}

\begin{figure}
\centering
\includegraphics[width=0.65\textwidth]{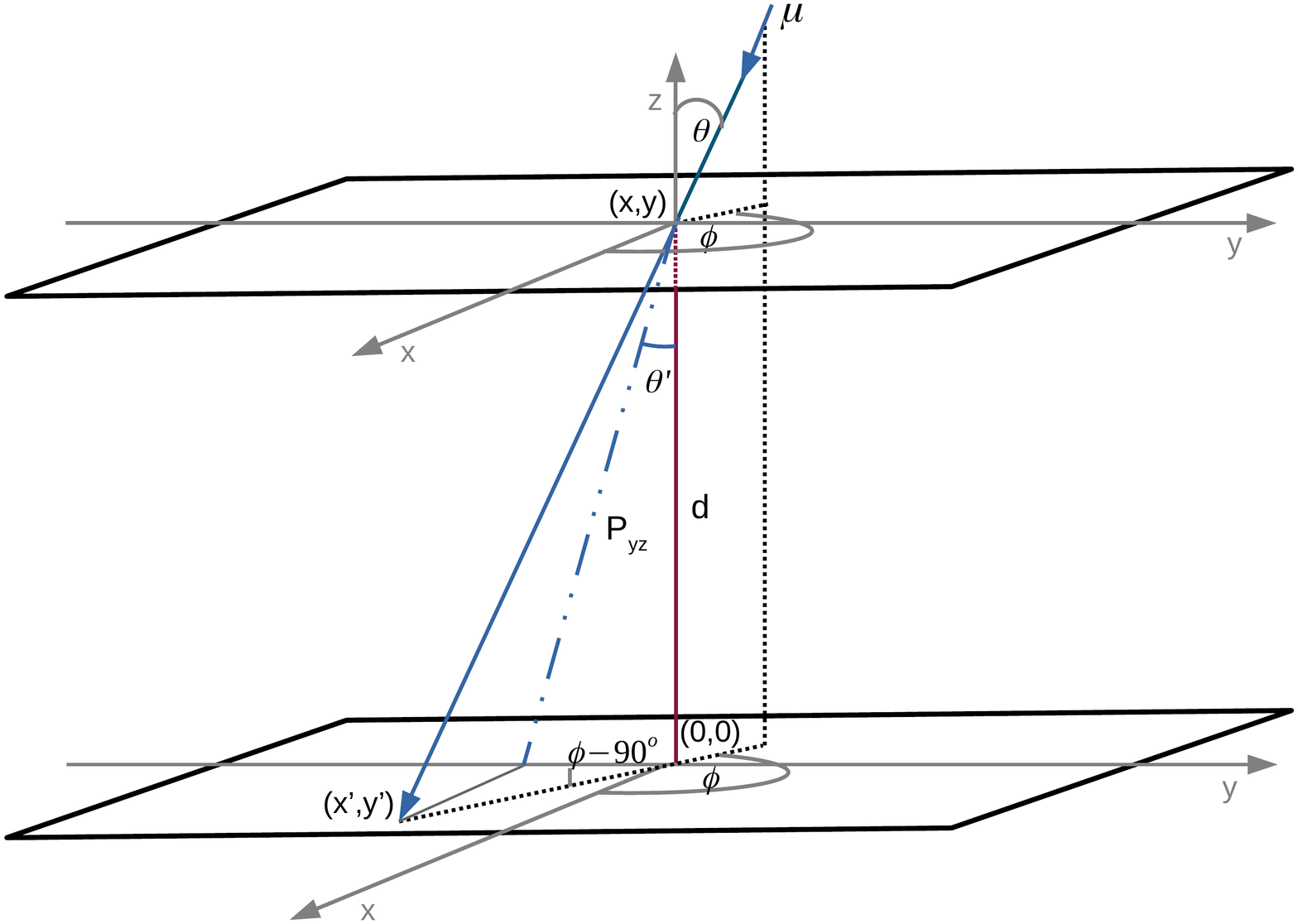}
\caption{Picture used for the first estimate of acceptance using lower plane of top
  detector and top plane of bottom detector.}
\label{Fig15SimDetGeo}
\end{figure}
To obtain the accepted muon distribution in the detector setup, Monte Carlo simulations are
performed. 
The muon events ($N_m=7000000$) are uniformly generated at the lower face of top detector
using Monte Carlo generator uniformly in $\phi$ and as per $\theta$ given by
\begin{equation}
  I(\theta) = I_0 \,\, \cos^n\theta
\label{cost}
\end{equation}
The value of exponent $n\sim 2.1$ is obtained by fitting the simulated and accepted
distribution with the experimentally measured distribution. 
We also generate energy deposition in each scintillator using the measured energy shape
for vertical muons using Langau function (parameters are given in Table~\ref{table5}).
 These energy depositions are scaled as per the pathlengths decided by the position of 
 event and the generated angles $\theta$ and $\phi$.
 Figure~\ref{Fig15SimDetGeo} shows the picture used for the first estimate of acceptance
using lower plane of top detector and top plane of bottom detector separated
by a distance $d$. Let us define that the z-axis is perpendicular to the plane of
the detectors and the x- and y-axis lie in the horizontal plane of the detectors.
Now assume that a muon traveling at a certain zenith angle $\theta$ and an
azimuthal angle $\phi$ hits the top detector
bar at the coordinates $(x,y)$, then it travels and hits the bottom bar at position 
$(x',y')$ which can be calculated as 
\begin{equation}
(x',y') = (x - d \, \tan\theta \, \cos\phi, \,\,\, y - d\,\tan\theta\,\sin\phi).
\label{rd1}
\end{equation}
If the point $(x',y')$ falls within the area of the lower detector it is considered in the
next step.
Out of these selected events, only those muons are accepted which travel pathlengths
to have minimum energy
deposition (corresponding to the energy cut) in both top and bottom detectors.
 This way we get an accepted $\theta$ distribution in the detector.
Since we measure $\theta$ only in 2 dimensions (yz plane) but having a finite $\phi$ coverage,
the measured $\theta'$ ($\phi$ corrected) will be given by
\begin{equation}
\theta' = \cos^{-1}\Bigg(\frac{d}{P_{yz}}\Bigg)
\label{rd2}
\end{equation}
where $P_{yz}$ is the projected track length on the yz plane.


\begin{figure}
\centering
\includegraphics[width=0.90\textwidth]{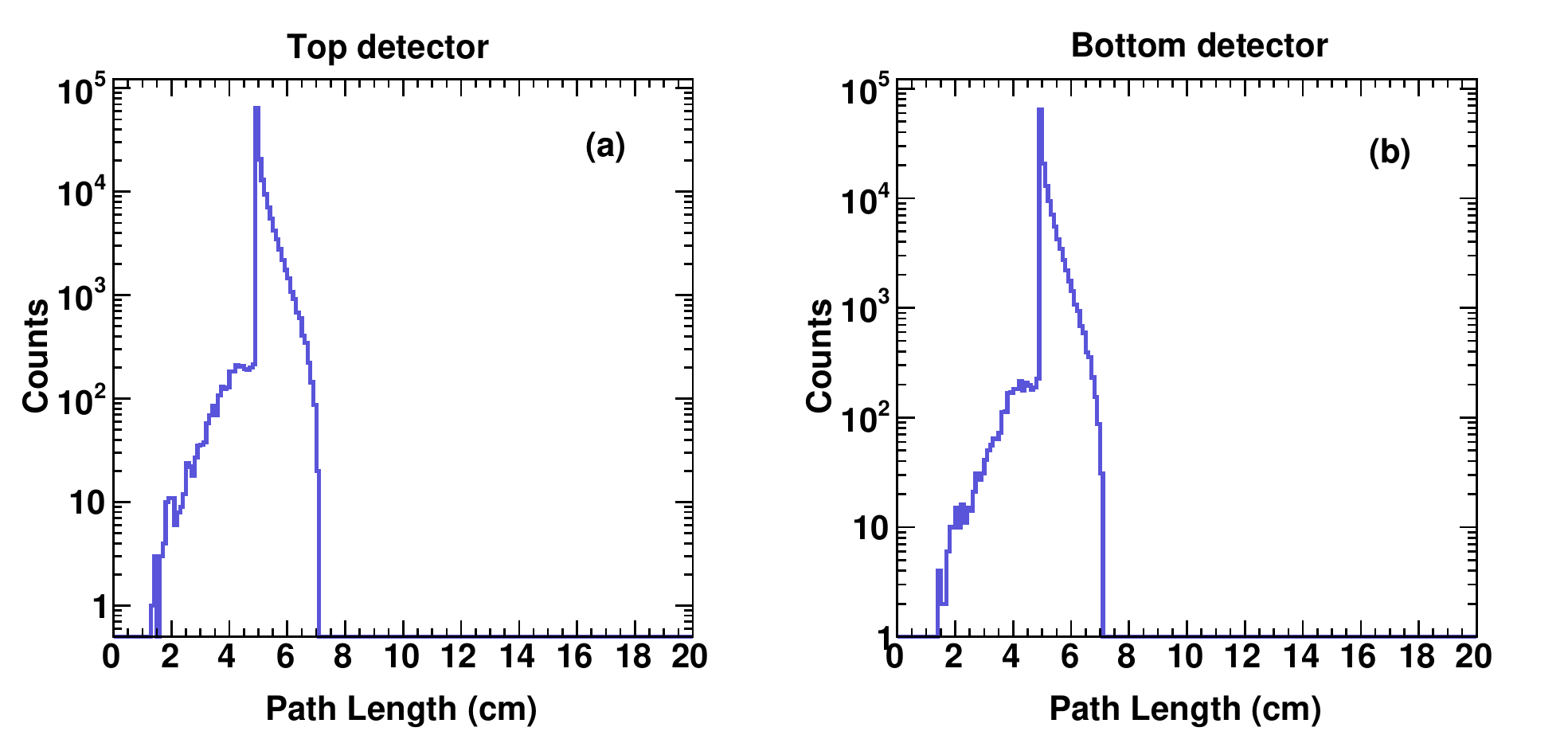}
\caption{The distributions of pathlength traversed by the accepted muons in
  the  (a) Top and (b) Bottom detectors for energy deposition above 8.07 MeV
  in the Square geometry $G_{14}$.}
\label{Fig16SimPath}
\end{figure}

\begin{figure}
\centering
\includegraphics[width=0.80\textwidth]{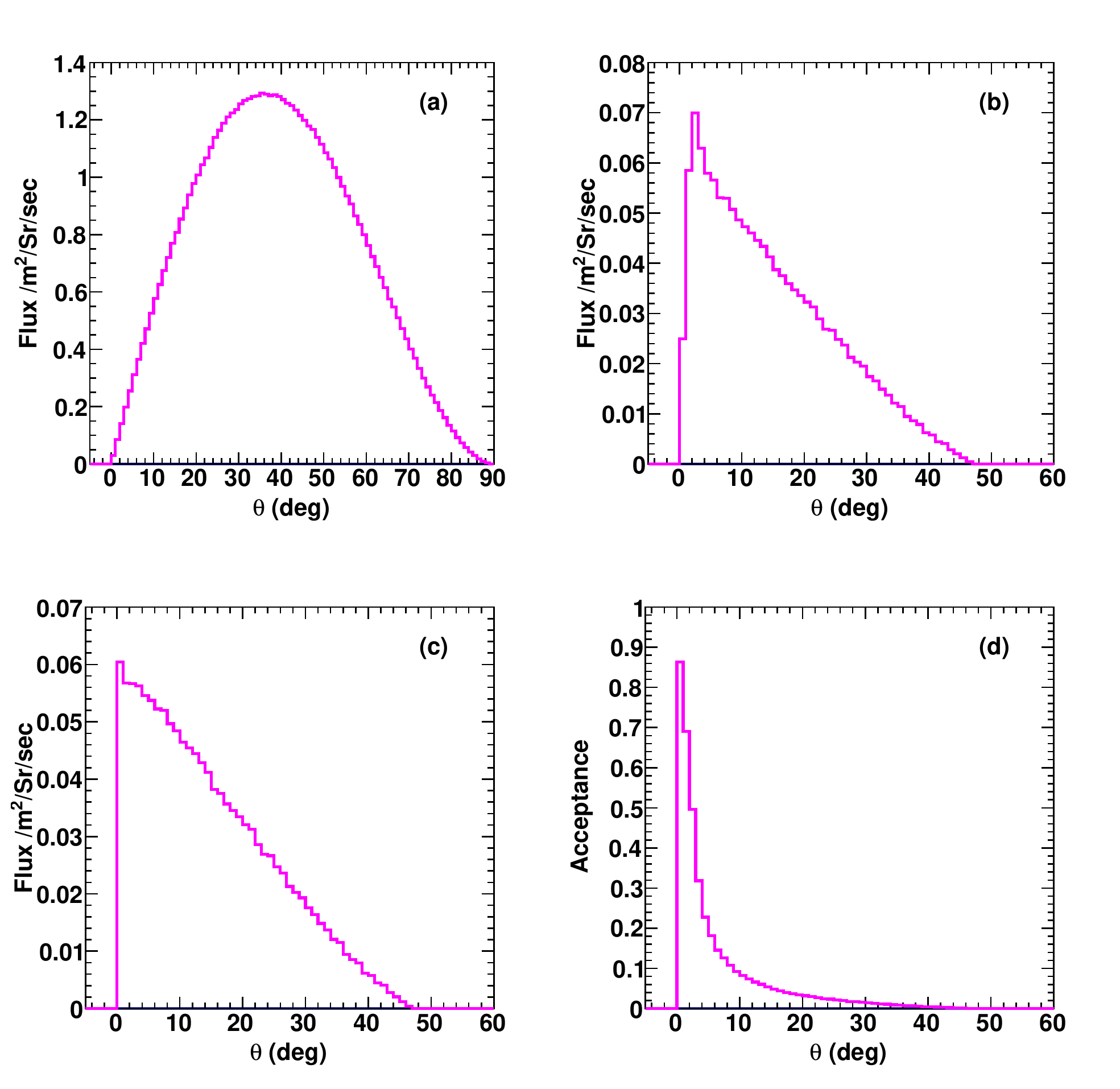}
\caption{(a) The generated zenith angle $\theta$ distribution
on the top detector using $n=1.86$ and normalized to $I_{\circ}=65.38/m^2/sr/s$. 
  (b) Accepted $\theta$ distribution (c) Accepted $\theta$ distribution ($\phi$ corrected).
  (d) Acceptance $\alpha$ which is Accepted/Generated, in the Square geometry $G_{14}$.}
\label{Fig17SimTheta}
\end{figure}

Figure~\ref{Fig16SimPath} shows the distributions of pathlength traversed
by the accepted muons in the (a) Top and (b) Bottom detectors for energy
deposition above 8.07 MeV in the Square $G_{14}$ geometry. It is clear from both
figures that for energy deposition less than 8.07 MeV, muons with smaller
pathlength will be rejected, such muons actually pass through the side faces of
the detectors.

 Figure~\ref{Fig17SimTheta}(a) shows the generated zenith angle distribution
on the top detector using $n=1.86$ and total events = 7000000 normalized to $I_{\circ}=65.38/m^2/sr/s$. 
Figure~\ref{Fig17SimTheta}(b) shows the Accepted $\theta$ distribution.
Figure~\ref{Fig17SimTheta}(c) shows the Accepted $\theta$ distribution 
($\phi$ corrected) as mentioned in Eq.~\ref{rd2}.
Figure~\ref{Fig17SimTheta}(d) shows the acceptance which is the
accepted/generated (counts in (b)/counts in (a)) in the Square geometry $G_{14}$ after muon
minimum energy deposition 8.07 MeV in  the top and bottom detectors.

The acceptance $\alpha$ is then obtained by ratio of such accepted events to the generated
events as 

\begin{equation}
{\rm \alpha} = \frac{\rm Accepted\,\,events}{\rm Generated\,\,events} 
\label{eqacc}
\end{equation}

\section{Muon flux measurements using different geometries}
  In this section, we present the muon flux measurements using the geometries
given in section 2. Our most optimized geometries are the Square geometries 
$G_{14}$ and $G_{13}$. 
 These are called Square geometries since the distance between top and bottom detectors is 
 about 95 cm giving a shape of Square looking from a side (yz plane).
 The inter detector distances
 are kept almost equal and are nearly 3 times larger than the distances corresponding to
 time resolution of the detectors.
 This allows inter detector timings to be used to decide track quality.
 Our setup is capable of distinguishing between upgoing track and 
 downgoing tracks. The position resolution involving 4 bars is 1.5 cm. 
 The track angle resolution for this geometry is 1 degree for higher 
 angles which increases to 2 degrees for small angles.

\begin{figure}
\centering
\includegraphics[width=0.95\textwidth]{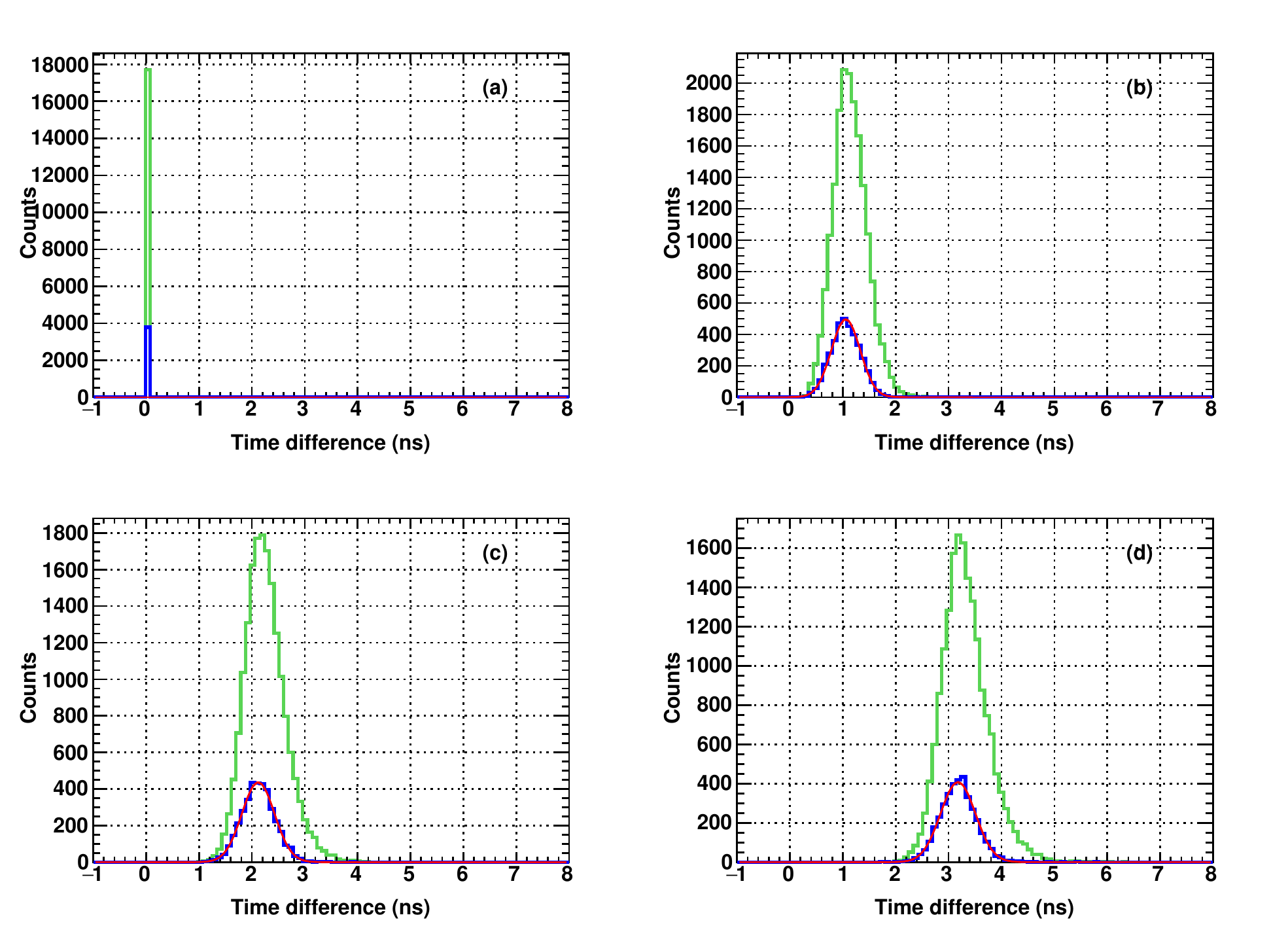}
\caption{Time taken by a muon track starting from 
top detector (3) in the Square geometry $G_{14}$ for (a) Detector 3 (b) Detector 1
(c) Detector 0 and (d) Detector 2. The distances of the three detectors
from the top detector corresponds to 31.7 cm, 63.2 cm and 95.0 cm. 
Green histogram is for muon events in all 4 detectors with final cuts 
and blue histogram is for vertical muon events (tolerance is $\pm 5^\circ$) 
in all 4 detectors with final cuts. This histogram is fitted with gaussian function. }
\label{Fig18AmongDetT}
\end{figure}

Figure~\ref{Fig18AmongDetT} shows the time taken by a muon track starting from 
top detector (3) in the Square geometry $G_{14}$ for (a) Detector 3 (b) Detector 1
(c) Detector 0 and (d) Detector 2. The distances of the three detectors
from the top detector corresponds to 31.7 cm, 63.2 cm and 95.0 cm. 
Green histogram is for muon events in all 4 detectors with final cuts 
and blue histogram is for vertical muon events (tolerance is $\pm 5^\circ$) 
in all 4 detectors with final cuts. This histogram is fitted with gaussian function.  
Green histogram is broader towards right side due to tracks corresponding
to larger zenith angles.

\begin{figure}
\centering
\includegraphics[width=0.7\textwidth, height=0.7\textwidth]{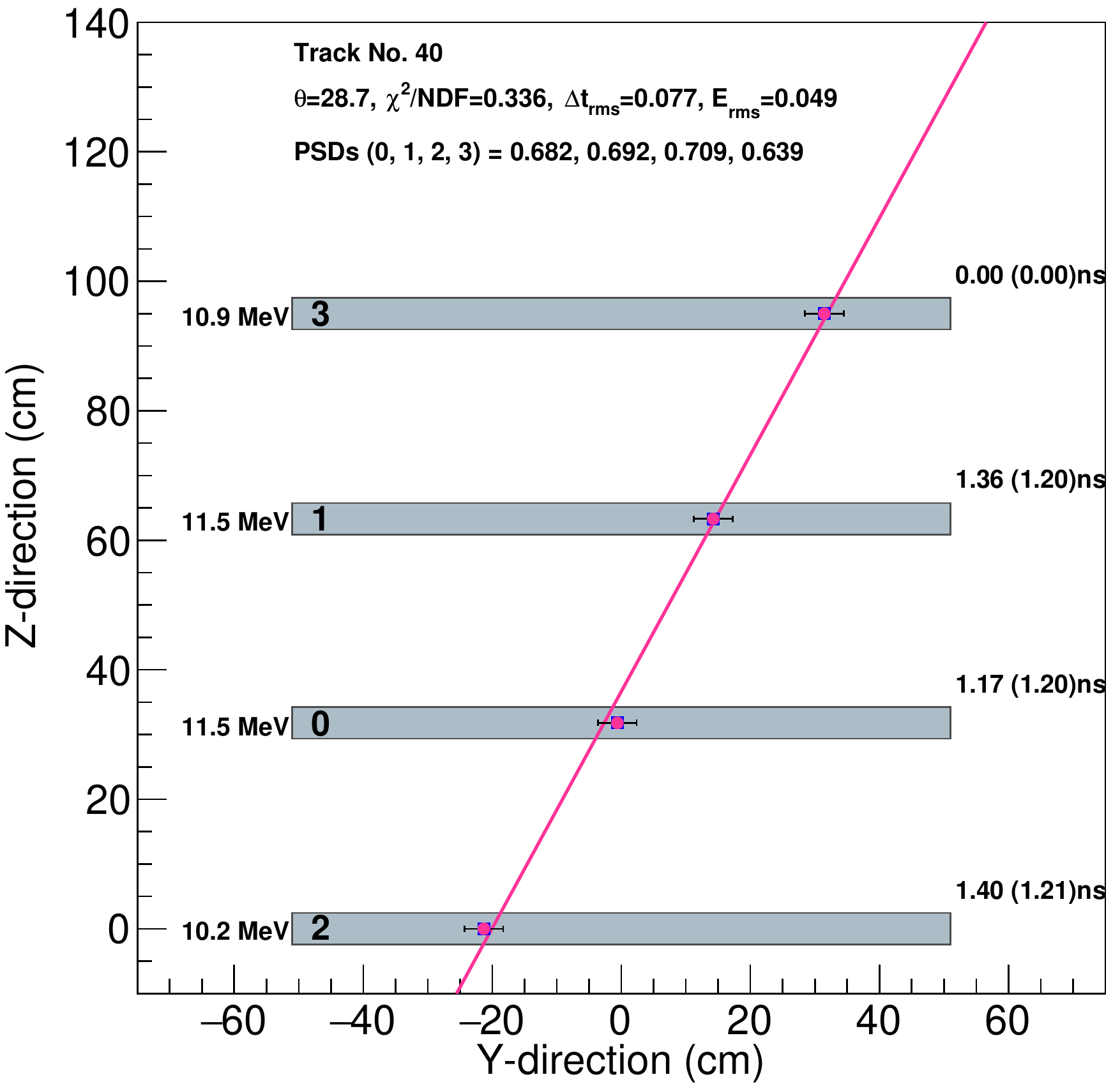}
\caption{A muon track reconstructed in the Square geometry $G_{14}$ along
  with quality parameters explained in the text. The timings written towards
  the right side of the detectors are the time taken by the muon track between the two detectors in
  nanoseconds.
  The time values in brackets are time taken by a light ray to traverse between the two detectors
  in the same direction as the track.
  The value of energy deposited in each detector is written towards the left side of each detector.
}
\label{Fig19trackG14}
\end{figure}

\begin{figure}
\centering
\includegraphics[width=7.5cm, height=7.5cm]{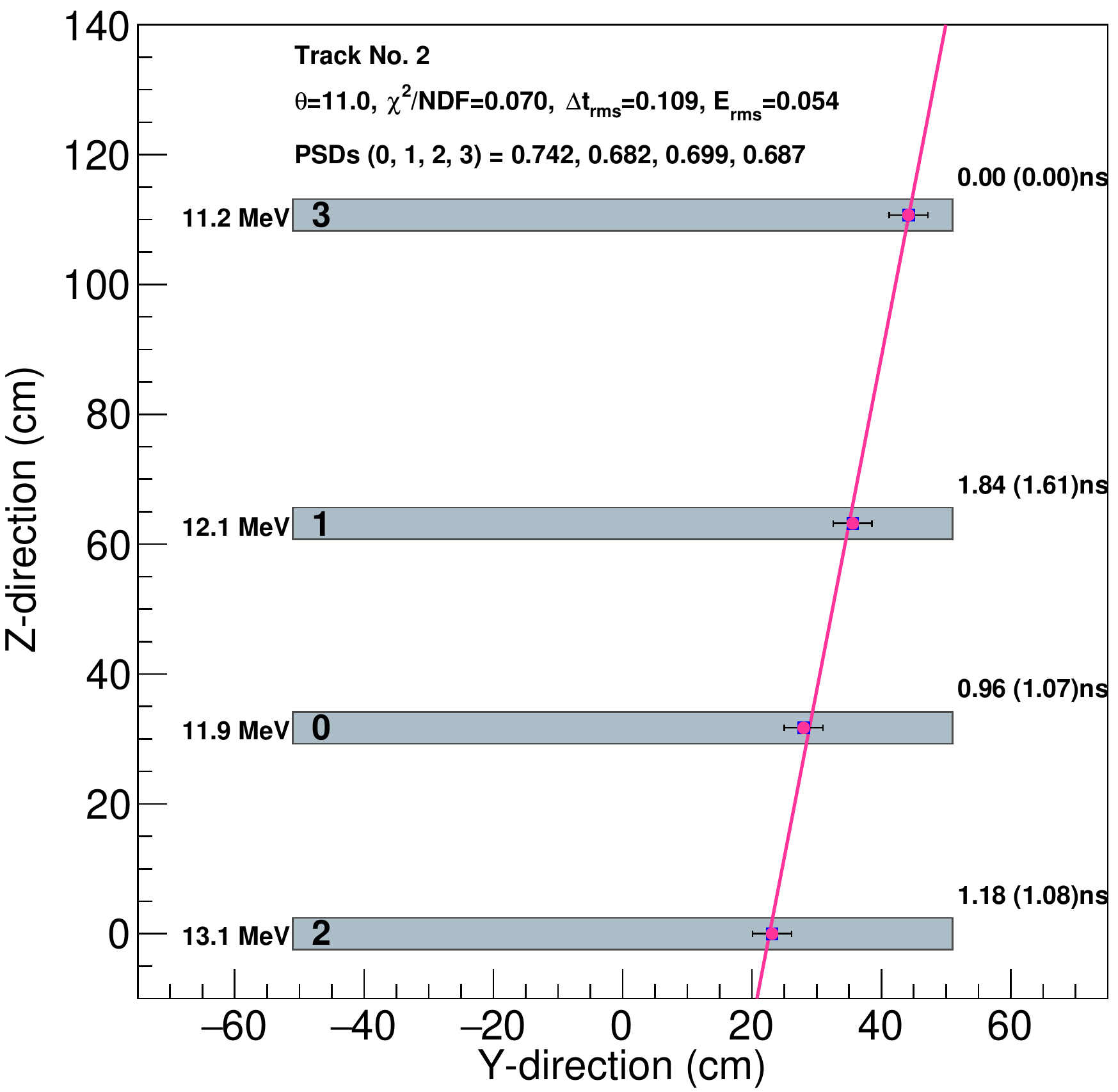}
\includegraphics[width=7.5cm, height=7.5cm]{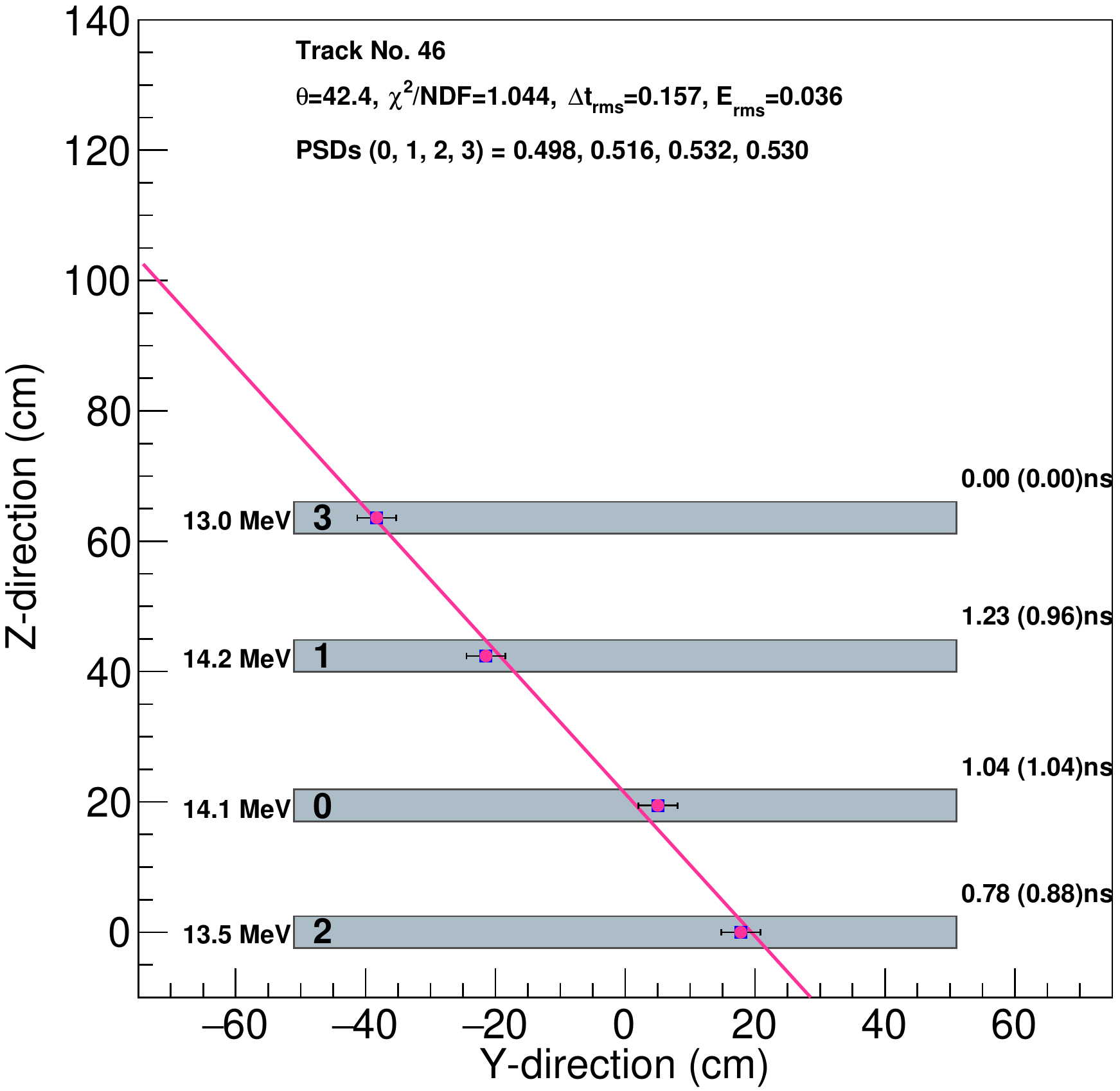}
\caption{A muon track reconstructed in (a) Large $G_{11}$ geometry and
  (b) Small $G_{17}$ geometry along with the track quality parameters explained in the text.}
\label{Fig20tracksG11}
\end{figure}

In the analysis, we can visualize the events track by track along with
the track parameters. This has been very useful in devising the selections on various
quality parameters.
Figure~\ref{Fig19trackG14} shows a muon track reconstructed in the Square
geometry $G_{14}$ along with quality parameters explained in the
previous section. The timing values written towards
the right side of the detector are the time (ns) taken by the muon track between the two detectors.
The times in brackets are time taken by a light ray to traverse
between the two detectors in same direction as the muon track.
The value of energy deposited by the track in each detector is written towards left side of each detector.

 We also use a Large Geometry $G_{11}$ corresponding to an increased distance of
the top detector from bottom. The Small Geometry $G_{17}$ is a squeezed setup
of four bars and the inter detector distances are kept equal. This geometry
can cover zenith angles upto 60 degrees but with some increase in angular resolution. 
Figure~\ref{Fig20tracksG11} shows a muon track reconstructed along with
the corresponding track quality parameters in the setups
(a) Large $G_{11}$ and (b) Small $G_{17}$.

\begin{figure}
\centering
\includegraphics[width=0.6\textwidth, height=0.6\textwidth]{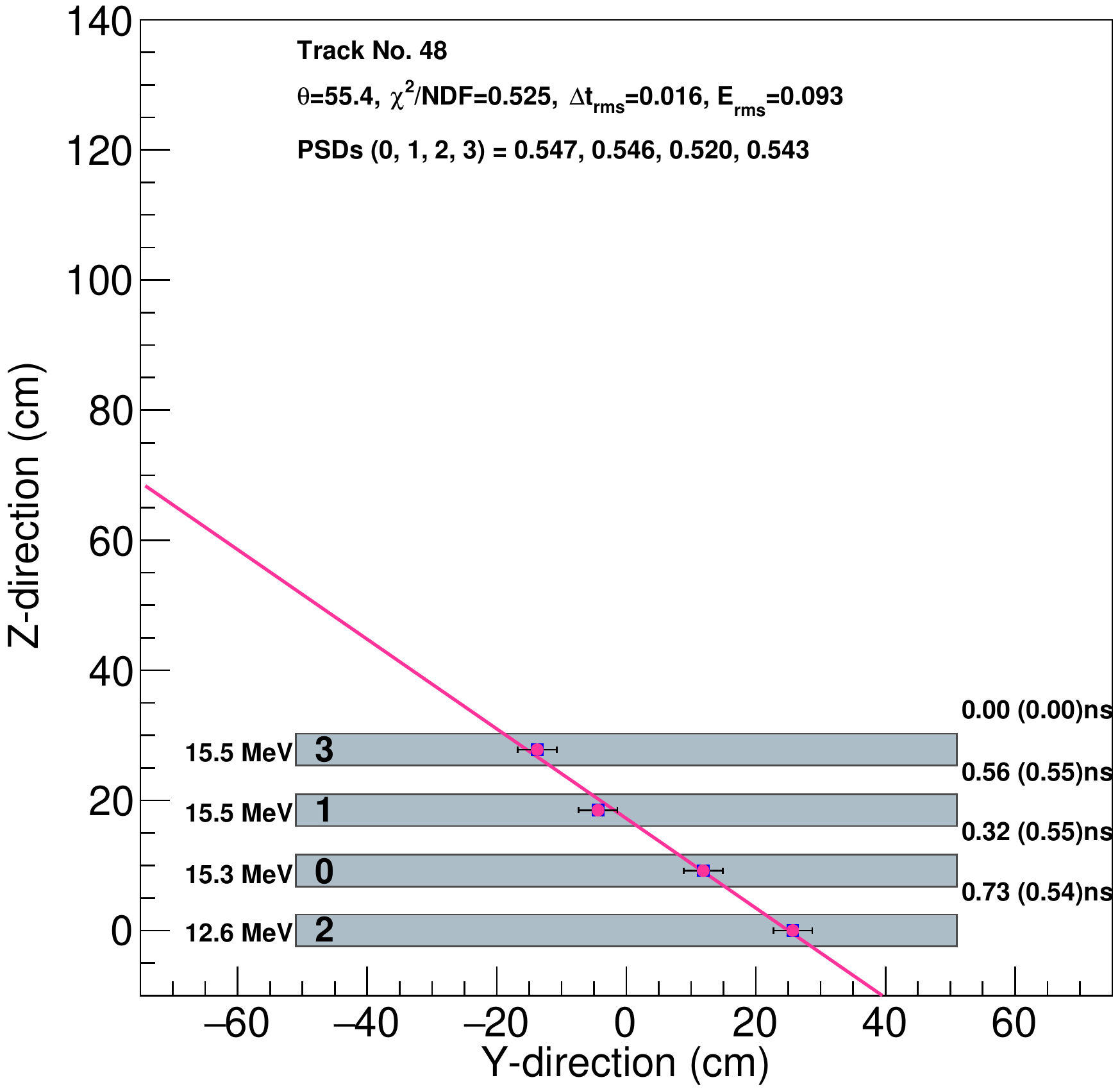}
\caption{A muon track reconstructed in the Close geometry $G_{16}$ along
  with the track quality parameters explained in the text.}
\label{Fig21tracksG16}
\end{figure}

Close geometries $G_{15}$ and $G_{16}$ are used to measure the efficiencies of the detectors
where the inter detector separations are kept very small. The angular resolution
of these geometries are very poor and thus will not give a good zenith angle distribution.
Figure~\ref{Fig21tracksG16} shows the muon track reconstructed in the Close geometry $G_{16}$ along 
with the track quality parameters.

\begin{figure}
\begin{center}
\includegraphics[width=0.8\textwidth, height=0.45\textwidth]{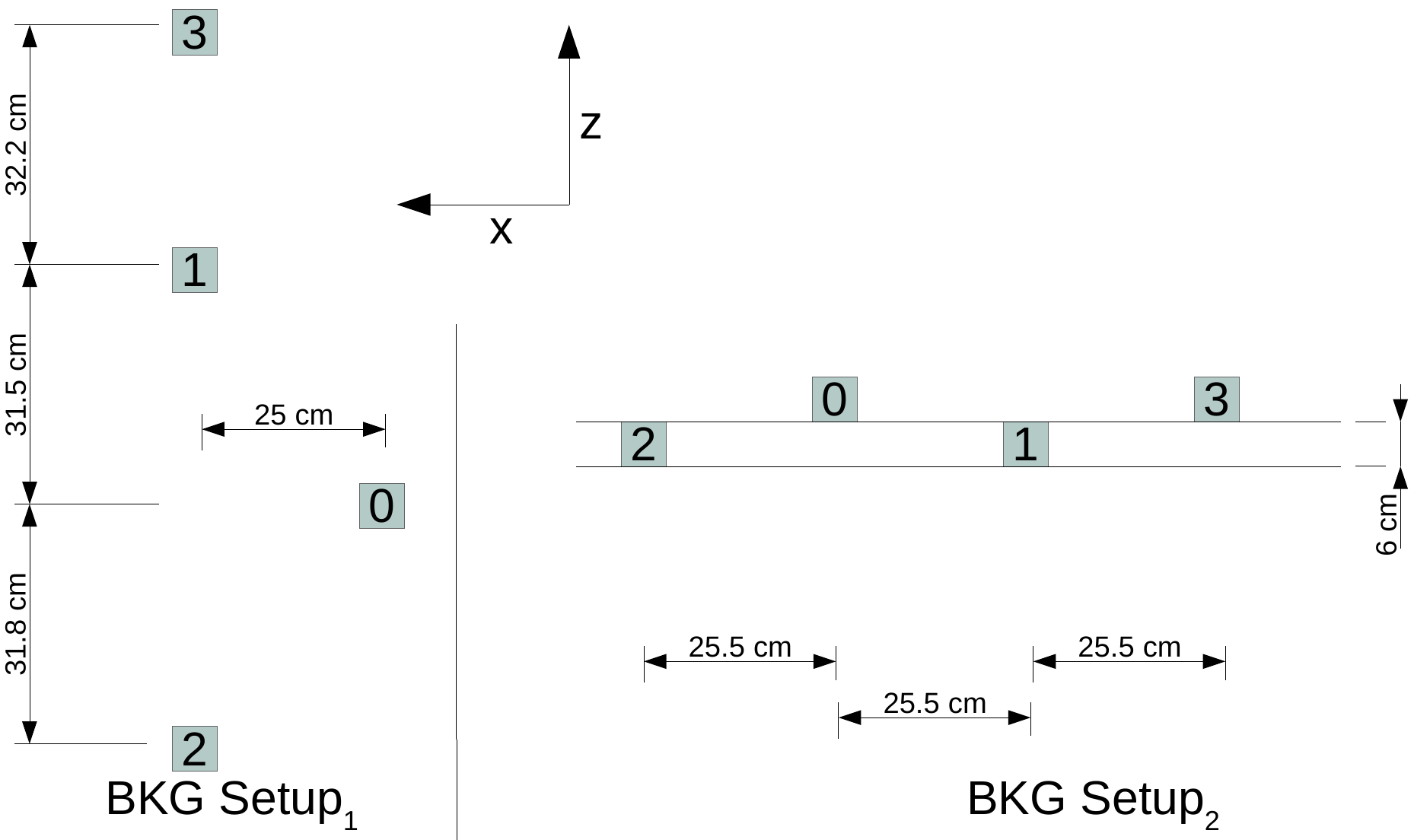}
\caption{Setups used to measure the uncorrelated background.}
\label{Fig22BKGsetup}
\end{center}
\end{figure}

To measure the backgrounds we use two kinds of setups. One in which we shift
one of the middle detectors in the Square geometry by a distance of 25 cm. In another setup,
we place the detectors horizontally such that alternating detectors are shifted
up by 6 cm. Figure~\ref{Fig22BKGsetup} shows these two background setups.

 The corrected flux distribution in the detector can be estimated from the observed
muon counts as
\begin{equation}
I(\theta) = \frac{N_{\rm meas}(\theta)}{A \times \epsilon \times \Omega \times t} \,,
\label{fluxm}
\end{equation}
where the $N_{\rm meas}(\theta)$ is the number of events registered in the setup during data
acquisition time $t$. Here, $A$ is the area of top scintillator surface which is equal to
0.056 $\times$ 1.02 $m^2$ and $\epsilon$ is the
efficiency of setup is given by

\begin{equation}
\epsilon = \epsilon_{E}\times\epsilon_{1} \times\epsilon_{3} \times\epsilon_{3},
\end{equation}
where $\epsilon_{E}$ is the energy deposition and PSD cut efficiency of all the detectors,
three or four used and $\epsilon_{1}$, $\epsilon_{2}$, $\epsilon_{3}$ are the efficiencies
of the cuts of three track quality parameters. 
Here, $\Omega$ is the solid angle (which covers $\theta$ and $\phi$ from $0-\pi/2$
and $0-2\pi$ respectively) given in terms of exponent $n$ of the zenith angle
distribution by
\begin{equation}
\Omega = \frac{2\pi}{n+1}.
\end{equation}
For a given $n$, thus the flux is obtained by integrating $I(\theta)$ in
Eq.~\ref{fluxm} and divide by $\alpha$.
But in the detailed analysis, we fit $I_0$ and $n$ using accepted simulated distribution
as in Figure~\ref{Fig17SimTheta}(c) on the measured distribution in Eq.~\ref{fluxm}.
Note that $n$ enters Eq.~\ref{fluxm} through $\Omega$.

Table~\ref{table8} gives the comparison of events measured using 4/3 liquid scintillator
bars in Square ($G_{13}$, $G_{14}$), Large ($G_{11}$), Small ($G_{17}$), and
Close geometries ($G_{15}$, $G_{16}$) along with background setups.
This table is made with keeping $n=2$ fixed and demonstrates the comparison of
effects of different cuts on different geometries. $G_{14-2}$ and $G_{14-3}$ are the setup of 3 
detectors in which Det$_2$ and Det$_3$ are removed respectively from Square geometry
$G_{14}$.
Here distance in 2nd column corresponds to the distance between top and bottom detectors.
We mention here that the acceptance for background setups corresponds to the
distance in Square setup $G_{14}$ although the distances in actual
background setups (Figure~\ref{Fig22BKGsetup}) are smaller.

We define $I_1$, $I_2$, $I_3$, $I_4$ as the integrated muon fluxes (corrected with  
efficiencies and acceptance $\alpha$ using) under 
4 finalised cuts conditions as described below.

\begin{enumerate}
\item $I_1$: Particle gets detected in coincidence within 7 ns (5 ns) time in all 4 detectors
  and with finalised energy deposition $E_{cut}$ and PSD cut. 
\item $I_2$: All track events with finalised $\chi^{2}/NDF$ cut on $I_1$. 
\item $I_3$: All track events with finalised $\Delta t_{rms}$ cut on $I_2$.
\item $I_4$: All track events with finalised $E_{rms}$ cut on $I_3$.
\end{enumerate}

\begin{table}[ht]
\begin{center}
\caption{ The comparison of events measured from 4/3 liquid scintillator
bars in Square ($G_{13}$, $G_{14}$), Large ($G_{11}$), Small ($G_{17}$), and
Close geometries ($G_{15}$, $G_{16}$) along with background setups.
$I_1$, $I_2$, $I_3$, $I_4$ are the muon fluxes under the 4 cuts conditions as
described in the text. Here $n=2$ is fixed. Raw counts correspond to $I_4$. 
}
\label{table8}
\resizebox{\columnwidth}{!}{%
\begin{tabular}{|c|c|c|c|c|c|c|c|c|c|} 
\hline
{\bf Geometry} & {\bf Distance} & {\bf Time} & {\bf Acceptance} & {\bf $I_1$} & {\bf $I_2$}& {\bf $I_3$}  & {\bf $I_4$ $\pm$ Stat.} & Counts\\ 
& (cm) & (s) & {\bf $\alpha$} & $/m^2/sr/s$ & $/m^2/sr/s$ & $/m^2/sr/s$ & $/m^2/sr/s$ &    \\ 
\hline
$G_{14}$ (3,1,0,2) & 95 & 113107 & 0.0213 & 73.02 & 69.95 & 68.15 & 66.02 $\pm$ 0.5 & 17703 \\
\hline
$G_{13}$ (1,3,2,0) & 95 & 135733 & 0.0213 & 74.00 & 70.67 & 68.45 & 66.67 $\pm$ 0.45 & 21657\\
\hline
$G_{11}$  (3,1,0,2) & 110.7 & 104319 & 0.0167 & 74.27 & 71.09 & 69.47 & 67.20 $\pm$ 0.59 & 13030 \\
\hline
$G_{17}$ (3,1,0,2)  & 63.6 & 61495 & 0.0381 & 72.42 & 70.17 & 67.19 & 65.04 $\pm$ 0.5 & 16825\\
\hline
$G_{14-2}$  (3,1,0) & 63.2 & 113107 & 0.0384 & 75.01 & 71.18 & 69.41 & 66.21 $\pm$ 0.37 & 31726\\
\hline
$G_{14-3}$  (1,0,2)  & 63.3 & 113107 & 0.0384 & 75.29 & 70.76 & 69.32 & 66.85 $\pm$ 0.37 & 31866\\
\hline
\multicolumn{9}{|c|}{\bf Efficiency Setups} \\
\hline
$G_{15}$ (3,1,0,2) & 27.8 & 25459 & 0.1043 & 68.78 & 67.5 & 66.69 & 65.08 $\pm$ 0.47 & 19539\\
\hline
$G_{16}$ (3,1,0,2) & 27.8 & 27169 & 0.1042 & 69.42 & 68.32 & 66.55 & 65.00 $\pm$ 0.45 & 20574\\
\hline
\multicolumn{9}{|c|}{\bf Background Setups} \\
\hline
BKG Setup1 & 95 & 16051 & 0.0213 & 1.64 & 0.15 & 0.05 & 0.05 $\pm$ 0.04 & 2\\
(3,1,0,2) & & & & & & & &   \\
\hline
BKG Setup2 & 95 & 39149 & 0.0295 & 1.08 & 0.06 & 0.05 & 0.04 $\pm$ 0.02 & 4\\
(3,1,0,2)  & & & & & & & &   \\
\hline
BKG Setup2-2 & 63.2 & 39149 & 0.0384 & 2.24 & 0.60 & 0.37 & 0.25 $\pm$ 0.04 & 42\\
(3,1,0)  & & & & & & & &   \\
\hline
BKG Setup2-3 & 63.2 & 39149 & 0.0384 & 2.51 & 0.69 & 0.29 & 0.19 $\pm$ 0.03 & 31\\
(1,0,2)  & & & & & & & &   \\
\hline
\end{tabular}
}
\end{center}
\end{table}

One can note from Table~\ref{table8}, the effectiveness of the tracking cuts
by comparing the values of numbers corresponding to four fluxes. The
values of $I_4$ for all geometries duly corrected for acceptance and efficiencies
are consistent with each other shows robustness of our analysis method.
The table also shows that the measurements with only 3 detectors are reasonably good
if we use all the track quality cuts with the 3 bars.
We also give two setups to measure uncorrelated backgrounds which shows that
uncorrelated background is negligible for our measurements for both 4 and
3 detectors.
 The actual flux is obtained by fitting the measured zenith angle distribution with
simulated distribution using $I_0$ and $n$ as free parameters in the next section.

\section{Results with systematic uncertainties and discussions}

As we have stated that the actual flux is obtained by fitting the measured zenith angle
distribution with
simulated distribution using $I_0$ and $n$ as free parameters. In the following
we give the measured zenith angle distributions along with simulated distributions
for all setups with best fit values of $I_0$ and $n$ given in Table~\ref{table9}. 

\begin{figure}
\centering
\includegraphics[width=0.45\textwidth]{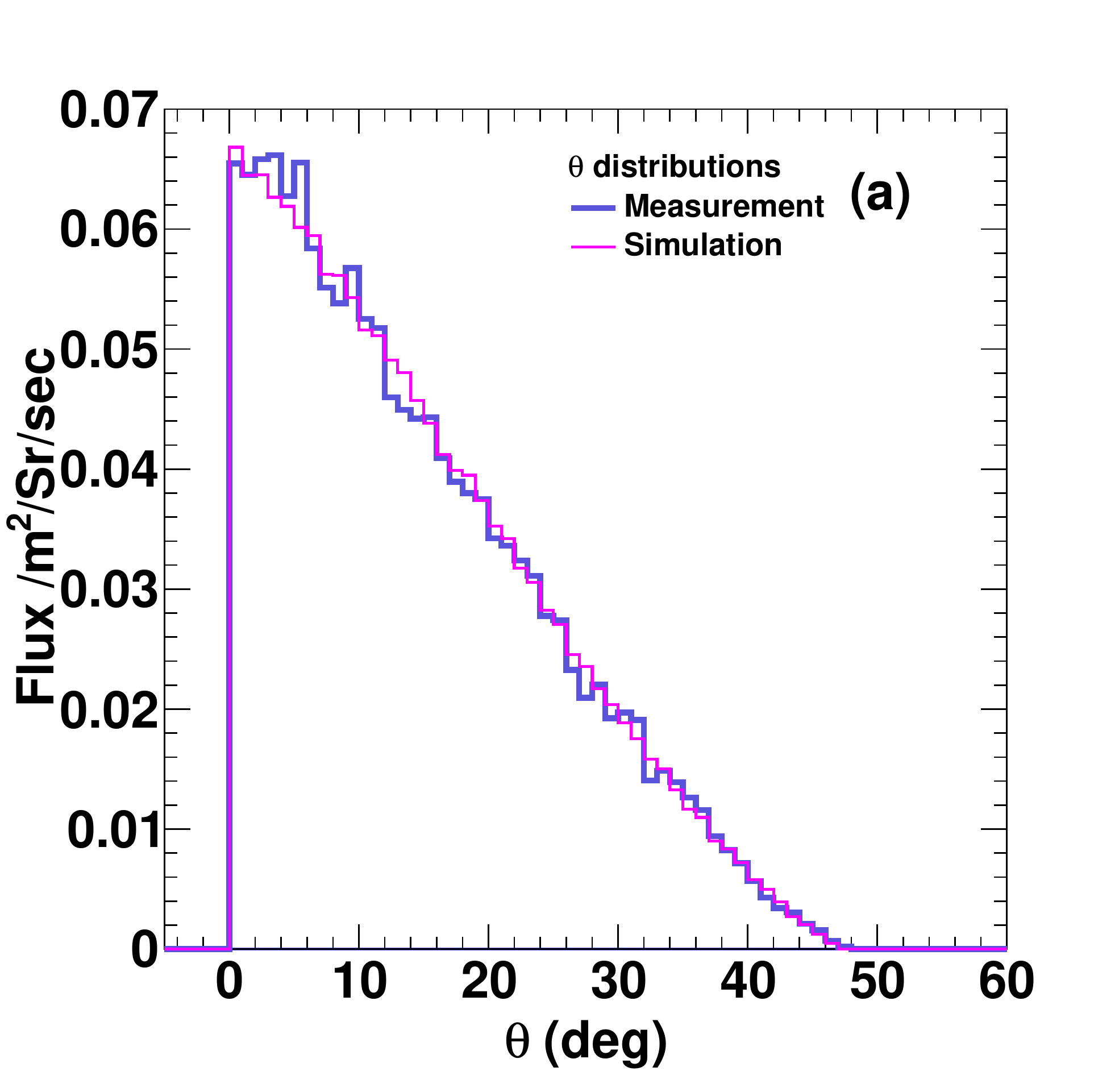}
\includegraphics[width=0.45\textwidth]{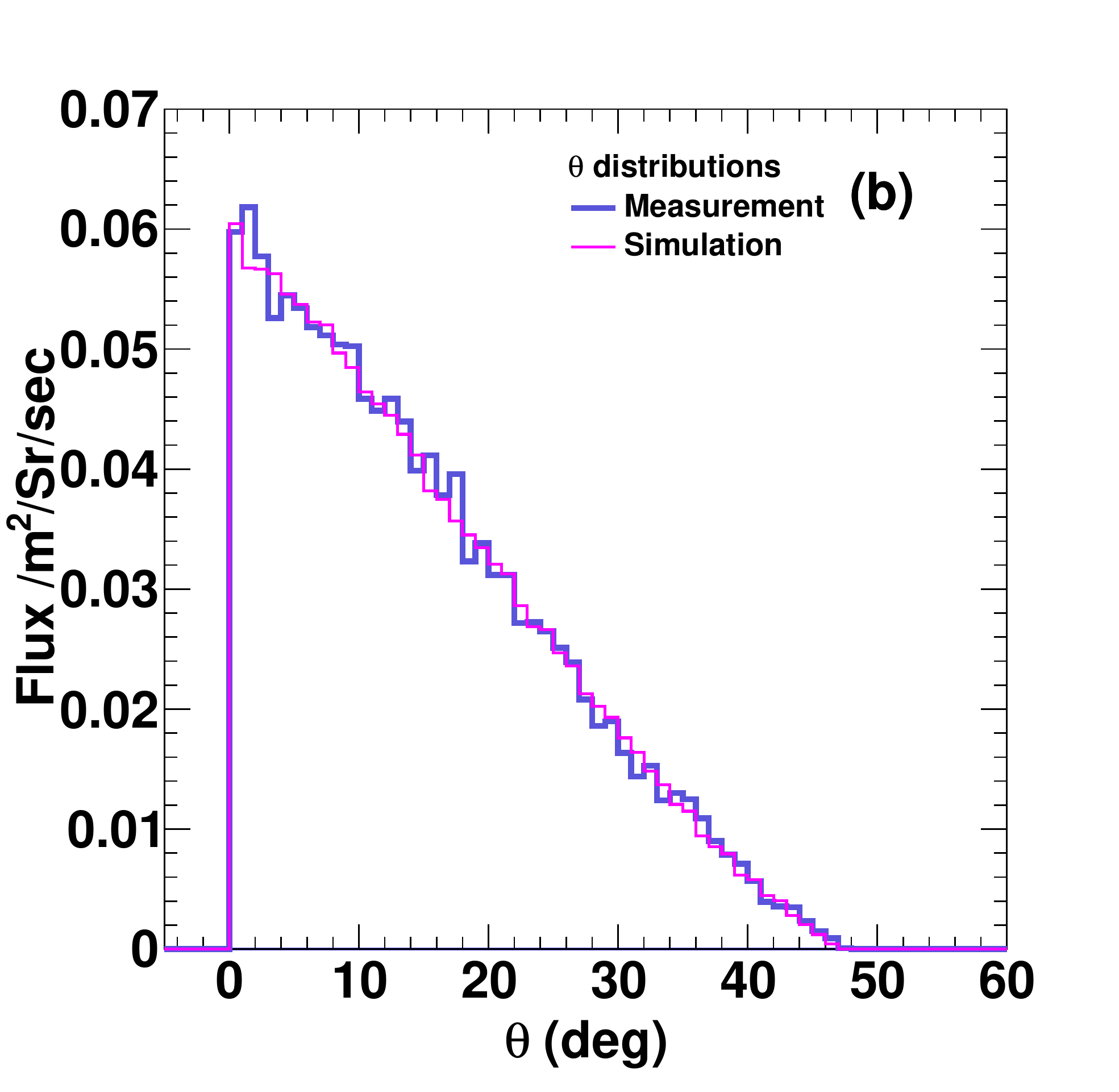}
\includegraphics[width=0.45\textwidth]{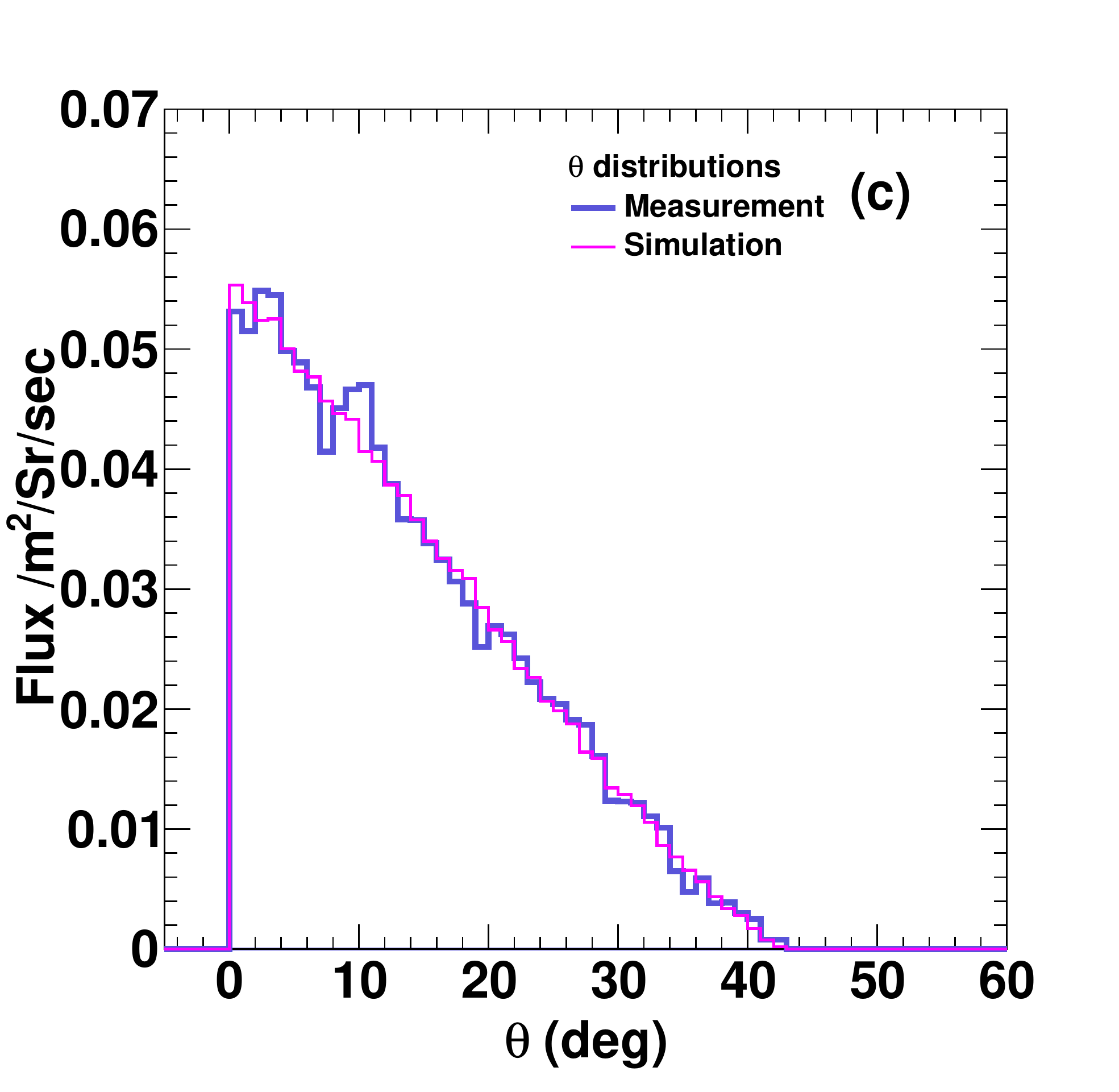}
\includegraphics[width=0.45\textwidth]{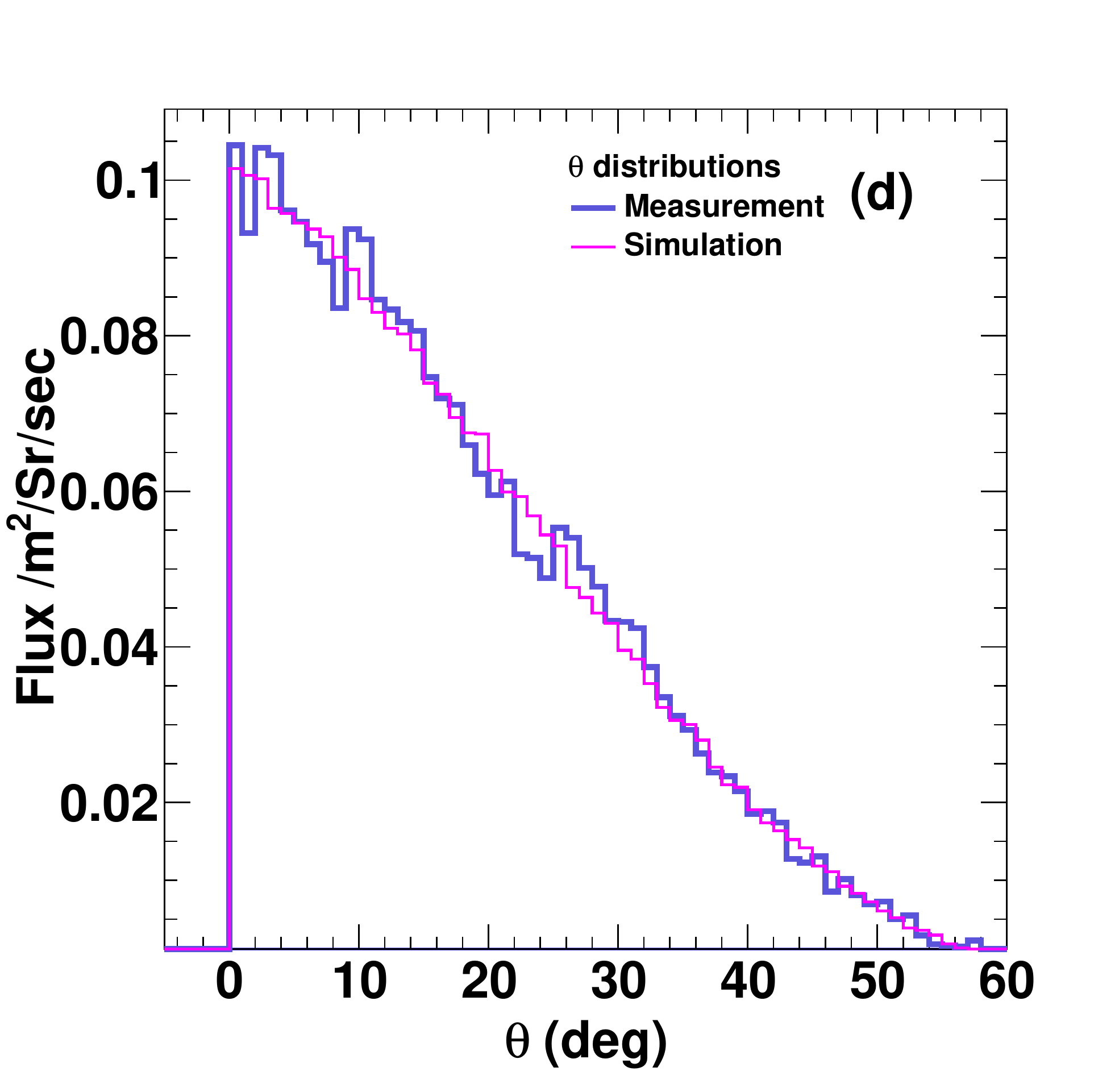}
\includegraphics[width=0.45\textwidth]{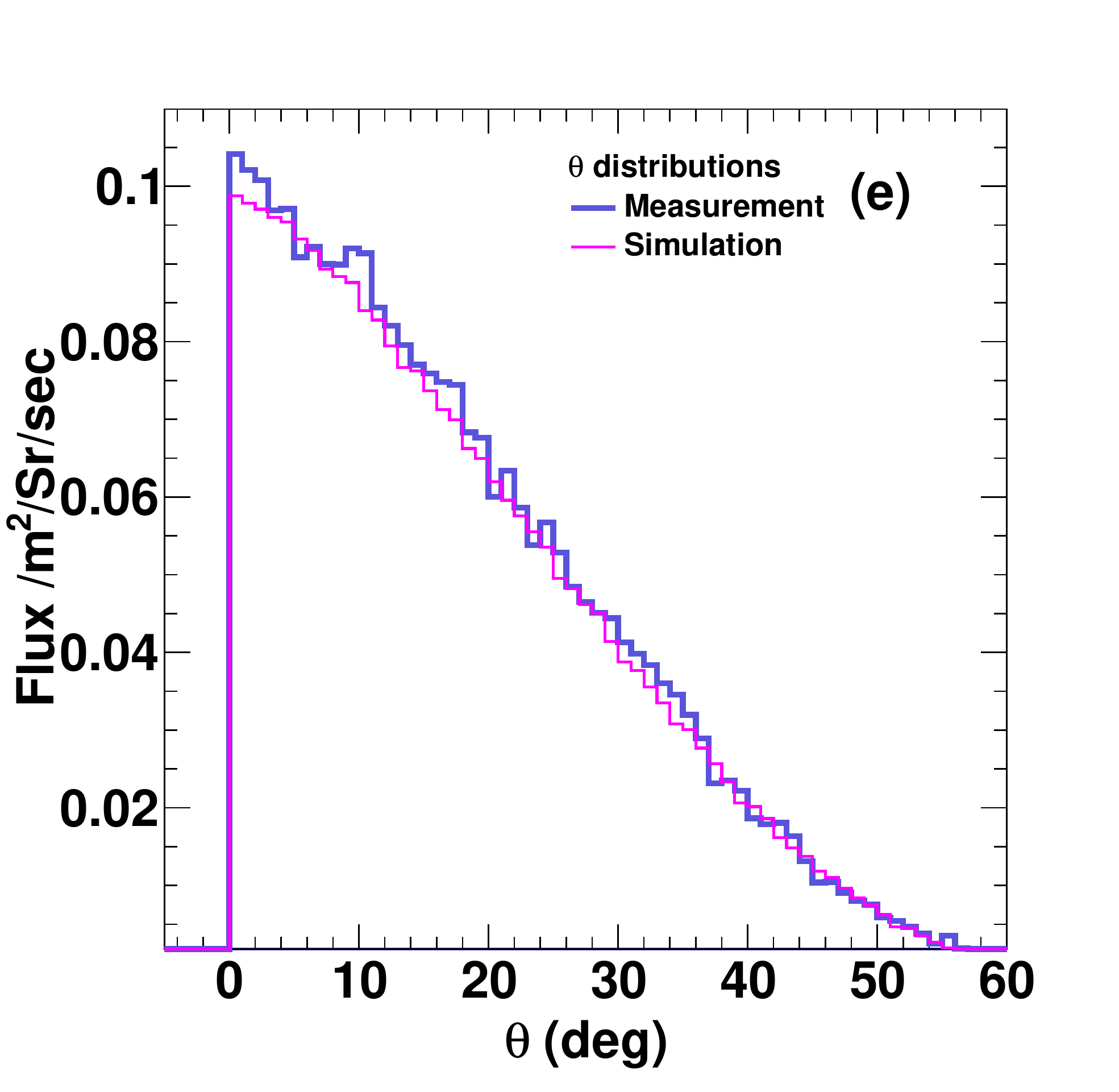} 
\includegraphics[width=0.45\textwidth]{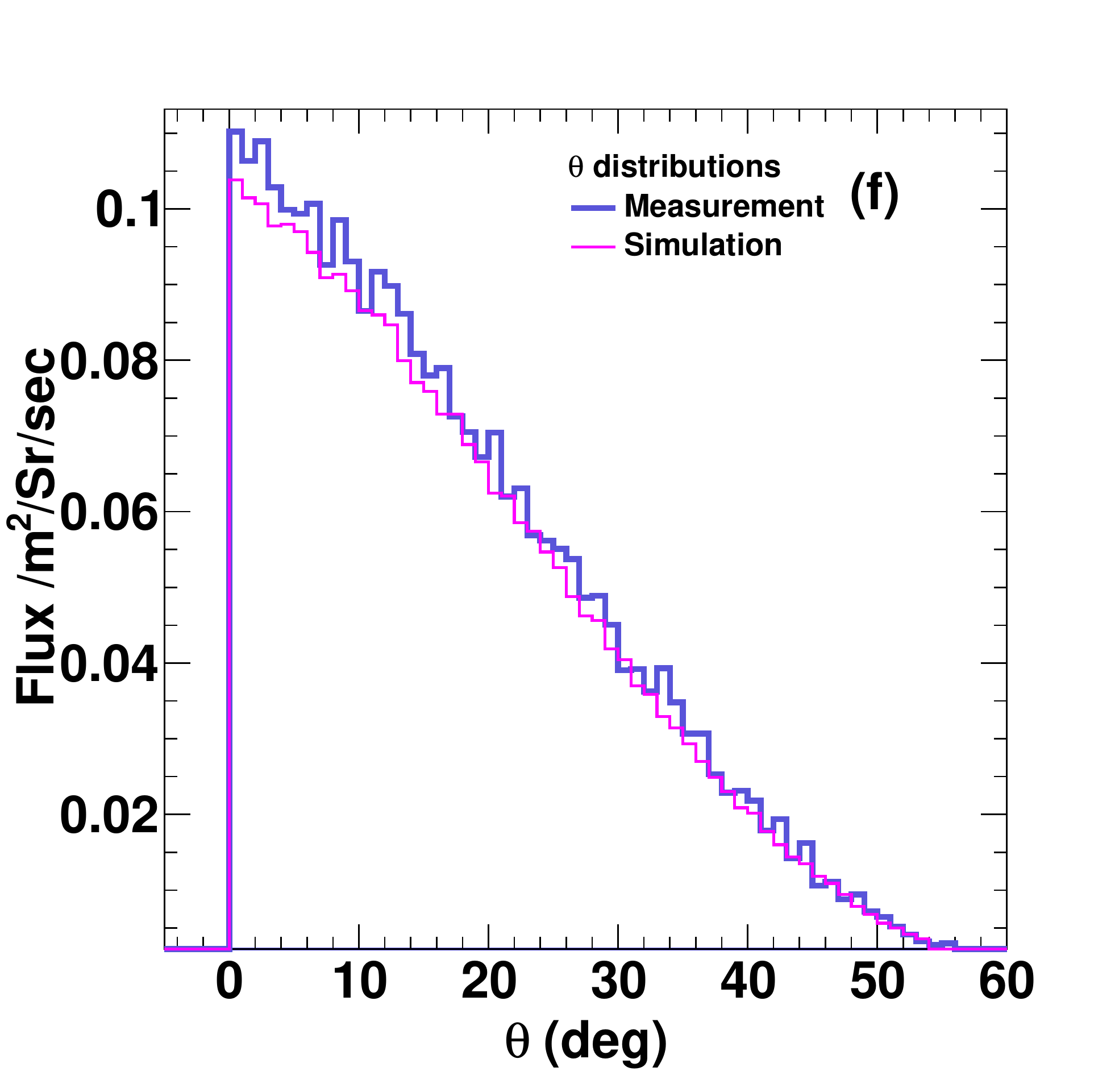} 
\caption{The measured and simulated zenith angle distribution of cosmic muons in geometries
  (a) Square $G_{13}$, (b) Square $G_{14}$, (c) Large $G_{11}$, (d) Small $G_{17}$ 
  (e) 3 Detector geometry $G_{14-2}$  and (f) 3 Detector geometry $G_{14-3}$ for
  best fit values of $I_0$ and $n$.}
\label{Fig23SimTheta13}
\end{figure}

\begin{figure}
\centering
\includegraphics[width=0.48\textwidth]{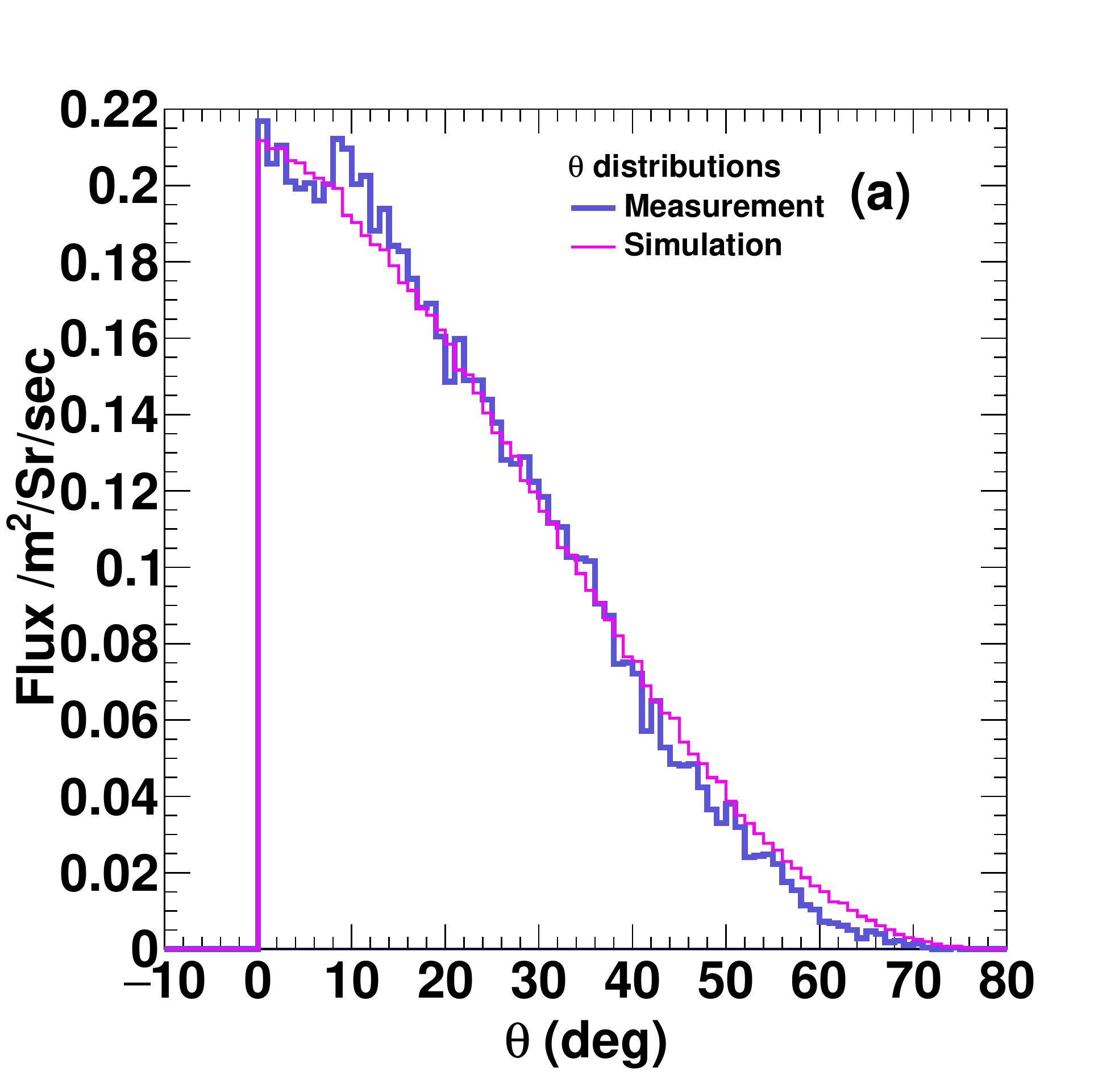}
\includegraphics[width=0.48\textwidth]{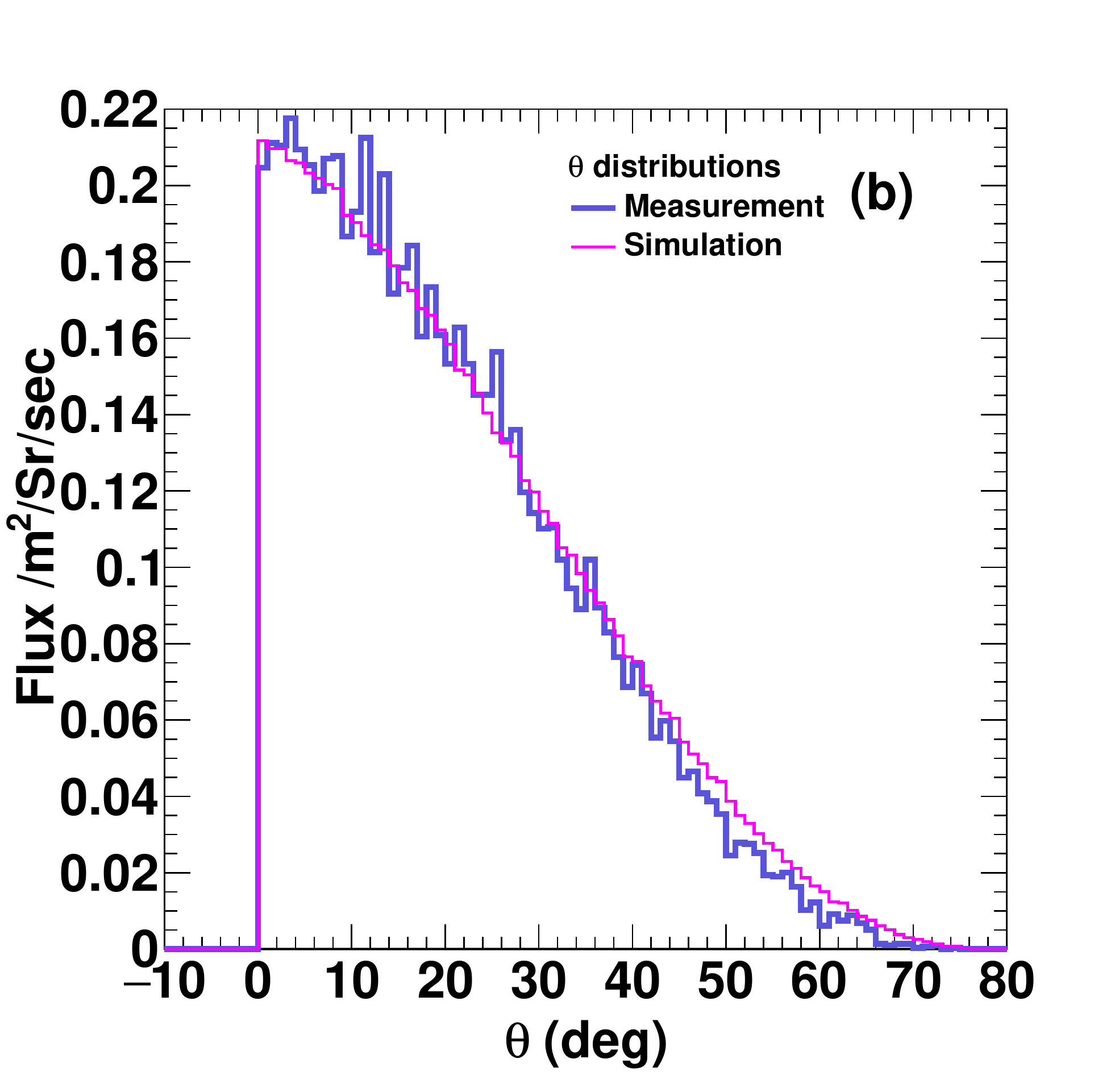}
\caption{The measured and simulated zenith angle distribution of cosmic muons in Close
  geometries (a) $G_{15}$ and (b) $G_{16}$ for best fit values of $I_0$ and $n$}
\label{Fig24SimTheta15}
\end{figure}

Figure~\ref{Fig23SimTheta13} shows the measured and simulated zenith angle distribution of 
cosmic muons in geometries (a) Square $G_{13}$, (b) Square $G_{14}$, (c) Large $G_{11}$, 
(d) Small $G_{17}$, (e) 3 Detectors $G_{14-2}$  and (f) 3 Detectors $G_{14-3}$ for best fit values
of $I_0$ and $n$. Figure~\ref{Fig24SimTheta15} shows the measured and simulated zenith angle
distribution of cosmic muons in Close geometries (a) $G_{15}$ and (b) $G_{16}$ for
fixed values of $I_0=66.70$ $/m^2/sr/s$ and $n=2.1$. The simulation and observed distributions are
not expected to have a good match for these geometries as the angular resolution becomes poor for smaller inter detector distances. 
The measurements presented in Figure~\ref{Fig23SimTheta13} and Figure~\ref{Fig24SimTheta15}
include efficiencies of finalised cuts on energy deposition, PSD and tracking quality. 
The value of these cuts and their respective efficiencies are given in Table~\ref{table6} and 
Table~\ref{table7}.

We use many geometries for obtaining systematic uncertainties as follows.

\begin{enumerate}

\item Setup-1: Flux from Square geometry $G_{14}$.
\item Setup-2: Flux from Square geometry $G_{13}$.
\item Setup-3: Flux from Large geometry $G_{11}$.
\item Setup-4: Flux from Small geometry $G_{17}$.
\item Setup-5: Flux from upper 3 detectors (3,1,0) in Square geometry $G_{14-2}$.
\item Setup-6: Flux from lower 3 detectors (1,0,2) in Square geometry $G_{14-3}$.
\item Setup-7: Flux from Square geometry $G_{14-t}$ changing the time-position calibration.
\item Setup-8: Flux from Square geometry $G_{13-t}$ changing the time-position calibration.
\item Setup-9: Flux from Square geometry $G_{14A-}$, with acceptance variation such that the distance
between top to bottom detectors is decreased by 0.5 cm.
\item Setup-10: Flux from Square geometry $G_{14A+}$, with acceptance change such that the distance
between top to bottom detectors is increased by 0.5 cm.


\end{enumerate}
The above variations are carefully planned to cover all possible uncertainties affecting
the measurements. The acceptance calculation and efficiency measurements are performed for each
geometry separately. 
Varying the distances among the detectors covers most of the systematics like
errors due to misalignments and
separations among detectors affecting acceptance of the detector which is usually a
major source of error. This also includes any variation coming from modeling the detector
in the simulations. 
 All the data has been taken during day time in the months of september, october and november 2021.
The three geometries with different distances also include the variation of measurement dates.
Interchanging detector positions include any bias due to any residual differences among the
detectors affecting angle distribution. We also measure after removing one of the detectors
to take into account any change due to the number of detectors involved. 
The time position relation affecting the value of $n$ has been taken into account. 

Table~\ref{table9} shows the values of $n$ and $I_0$ along with the fitting error
from different setups. The averaging and systematic error calculations
for flux and exponent for 10 Setups (N=20 including $\pm$ error on fitting)
are performed using the following equations.

\begin{eqnarray}
  I_{mean} & = &\frac{1}{N} \sum {I_i}, \\ 
  \Delta I_{rms} & = & \sqrt { \frac{1}{N} \sum_{i=1}^{N}
(I_i-I_{mean})^2 }.
\end{eqnarray}
Combining all these measurements the vertical muon flux measured is 
$66.70 \pm 0.36 (stat) \pm 1.50 (sys)$ $/m^2/sr/s$ with $n=2.10 \pm 0.05 (stat) \pm 0.25 (sys)$
in $\cos^n \theta$.

The measurement was carried out inside a single floor building with rooftop.
The muon momentum cutoff in our setup is calculated to be 255 MeV/c in simulation,
which includes the standard rooftop over the setup with 15 cm concrete and 10 cm bricks.

\begin{table}[ht]
\caption{The values of $I_0$ and $n$ for different measurements. 
All of them are used in calculating systematic error.}
\label{table9}
\begin{center}
\begin{tabular}{|l|c|c|c|c|c|c|c|c|c|} 
\hline
{\bf Setup}  & {\bf $I_{\circ}$ ($/m^2/sr/s$)} & {\bf $n$-value} & $\chi^{2}/NDF$ \\ 
& {\bf $I_{\circ}$  $\pm \,\sigma (fit)$} & {\bf $n \pm \sigma (fit)$} & (minimum)\\
\hline
1. Square $G_{14}$   & 65.375 $\pm$ 1.625 & 1.855  $\pm$ 0.125  & 0.910 \\
\hline
2. Square $G_{13}$   & 67.250 $\pm$ 1.150 & 2.120  $\pm$ 0.090  & 1.012 \\
\hline
3. Large $G_{11}$    & 66.875 $\pm$ 1.125 & 2.010  $\pm$ 0.200  & 1.395 \\
\hline
4. Small $G_{17}$    & 66.125 $\pm$ 0.875 & 2.285  $\pm$ 0.075 & 1.258 \\
\hline
5. $G_{14-2}$        & 67.650 $\pm$ 0.650 & 2.230  $\pm$ 0.100  & 1.071 \\
\hline
6. $G_{14-3}$        & 68.300 $\pm$ 0.800 & 2.345  $\pm$ 0.085  & 1.058 \\
\hline
7. Square $G_{14-t}$ & 67.125 $\pm$ 1.375 & 2.310  $\pm$ 0.100  & 1.070 \\
\hline
8. Square $G_{13-t}$ & 67.250 $\pm$ 0.750 & 2.195  $\pm$ 0.165  & 1.119 \\
\hline
9. Square $G_{14A-}$ & 65.030 $\pm$ 1.530 & 1.900  $\pm$ 0.260  & 0.956 \\
\hline
10. Square $G_{14A+}$& 66.000 $\pm$ 1.000 & 1.745 $\pm$ 0.135  & 0.979 \\
\hline
\end{tabular}
\end{center}

\end{table}

\begin{table}[ht]
\caption{Comparison of vertical muon flux with other experiments.}
\label{table10}
\begin{center}
\resizebox{\columnwidth}{!}{%
\begin{tabular}{|l|c|c|c|c|c|c|} 
\hline
{\bf References} & {\bf Geomag.} & {\bf $P_c$} & {\bf Altitude} & {\bf $P_{\mu}$} & {\bf $n$} & {\bf Flux  $I_{\circ}$} \\ 
 & {\bf Lat.} ($^{\circ}$N) & (GV) & (m) & (GeV/c) & & {\bf $(m^{-2}s^{-1}sr^{-1}$)} \\ 
\hline
Hayman et al.~\cite{Hayman1962}
 & 57.5 &   1.8  &   S.L.  &   $\geq 0.320$ & - & $76 \pm 0.6$\\
\hline
Greisen~\cite{Greisen1942}
 & 54 & 1.5 & 259 & $\geq$ 0.33  & 2.1 & 82 $\pm$ 1\\
\hline
Judge and Nash~\cite{Judge1965}
 & 53 & - & S.L. & $\geq$ 0.7  & 1.96 $\pm$ 0.22   & -\\
\hline
Crookes and Rastin~\cite{Crookes:1972xd} & 53 & 2.2 & 40 & $\geq$ 0.35  & 2.16 $\pm$ 0.01 & 91.3 $\pm$ 0.2\\
\hline
Barbouti and Rastin~\cite{Grieder2001}
 & 52 &   2.5  &   40  &  $\geq 0.438$ & - & $88.68\pm 1.15$ \\
\hline
Fukui et al.~\cite{Fukui1957}
 & 24 & 12.6 & S.L. & $\geq$ 0.34  & - & 73.5 $\pm$ 2\\
\hline
Gokhale~\cite{Grieder2001}
 & 19 & - &  124 & $\geq$ 0.27  & - & 75.5 $\pm$ 1\\
\hline
Karmakar et al.~\cite{Karmakar1973}
 & 16 & 15.0 & 122 & $\geq$ 0.353  & 2.2 &  89.9 $\pm$ 0.5\\
\hline
Sinha and Basu~\cite{Sinha1959}
 & 12 & 16.5 & 30& $\geq$ 0.27  & - & 73 $\pm$ 2\\
\hline
Allkofer et al.~\cite{pethuraj2017}
 & 9 & 14.1 &  S.L. & $\geq$ 0.32  & - & 72.5 $\pm$ 1\\
\hline
Pethuraj et al.~\cite{pethuraj2017}
 & 9.92 & 17.6 &  160 & $\geq$ 0.11  & 2.00 $\pm$ 0.04 $\pm$ 0.14   & 70.07 $\pm$ 0.02 $\pm$ 5.26  \\
\hline
Pal et al.~\cite{pal2012}
 & 19.07 & 16.38 & S.L.  & $\geq$ 0.280  & 2.15 $\pm$ 0.01   & 62.17 $\pm$ 0.05\\
\hline
Present data
 & 19.07 & 16.38 &  S.L.  & $\geq$ 0.255  & 2.10 $\pm$ 0.05 $\pm$ 0.25 & 66.70 $\pm$ 0.36 $\pm$ 1.50\\
\hline
\end{tabular}
}
\end{center}
\end{table}

Table~\ref{table10} shows the comparison of vertical muon flux values measured
by other experiments.
In general, the value of vertical muon flux increases with increasing the geomagnetic
latitude and altitudes although one can note from rows 2, 4 and 5 that there is large
variation among the measurements made at similar conditions.
Our value of muon flux is larger than the earlier measurement by Pal et al. \cite{pal2012}.
One reason is that our momentum cutoff is slightly smaller. Although exact systematic error
has not been estimated in their work, a later work using similar setup by
the same group \cite{pethuraj2017} gives a systematic error of 7.5$\%$ in flux
measurement. The systematic error in our flux measurement is 2.2$\%$.
Thus, they are consistent within the errors and the present value is more accurate.

\section{Summary and outlook}

In this work, we report measurement of muon flux and angular distributions using different
geometries of four one-meter long liquid scintillator bars.
We exploit energy and excellent timing of scintillators to construct two
dimensional tracks and hence angles of charged particles. Our analysis involves many
innovative methods.
  Position calibration of scintillator bars is performed using vertical cosmic muons constrained
by placing the four scintillators in cross positions. We also explore less time consuming
methods to connect the position with the measured time.
We propose three track quality parameters which are applied
to obtain a clean muon spectrum.
  We optimize the distances between the bars to have accurate zenith angle measurements.
For the most optimal distances among the bars, the zenith angle coverage is 0-45 degrees
which could be extended upto 60 degrees with reasonable precision. 
 With our improved analysis we demonstrate that a setup of 3 bars can also be used for
quick and precise measurements. 
The vertical muon flux measured is $66.70 \pm 0.36 (stat) \pm 1.50 (sys)$
$/m^2/sr/s$ with $n=2.10 \pm 0.05 (stat)\pm 0.25 (sys)$
in $\cos^n \theta$ at the location of Mumbai, India ($19^{\circ}$N, $72.9^{\circ}$E) at Sea 
level at a muon momentum above $255$ MeV/$c$. 
The muon flux has dependence on various factors, the most prominents are
latitude, altitude and a simple and portable setup like this is beneficial
for such measurements at various locations. The setup is capable of distinguishing
downgoing and upgoing tracks, protons and muon induced neutrons which will be
explored in the near future.

\acknowledgments

We would like to thank Dr. Vineet Kumar for initial help in the detector setup and
discussions. We thank Dr. Prakash Rout, Dr. Sandeep Joshi and Prof. Gobinda Majumder 
for many fruitful discussions and help during the measurements.

\end{document}